\documentclass[a4paper,11pt]{article}

\usepackage{jheppub}
\usepackage{lipsum}
\usepackage{xcolor}
\usepackage{etoolbox}
\usepackage{graphicx}
\usepackage{wrapfig}
\usepackage{amsmath}
\usepackage{amsfonts}
\usepackage{amssymb}
\usepackage{multirow}
\usepackage{slashed}
\usepackage{physics}
\usepackage{ulem}
\usepackage[colorlinks=true, linkcolor=blue, citecolor=blue, urlcolor=blue]{hyperref}

\newcommand{\pvec}{\vec{p}}
\newcommand{\qvec}{\vec{q}}
\newcommand{\xvec}{\vec{x}}
\newcommand{\yvec}{\vec{y}}
\newcommand{\zvec}{\vec{z}}

\DeclareRobustCommand{\Eq}[1]{Eq.~\eqref{eq:#1}}

\DeclareRobustCommand{\fig}[1]{Fig.~\ref{fig:#1}}

\DeclareRobustCommand{\sec}[1]{Sec.~\ref{sec:#1}}

\DeclareRobustCommand{\tb}[1]{Table~\ref{tb:#1}}

\DeclareRobustCommand{\refcite}[1]{Ref.~\cite{#1}}
\DeclareRobustCommand{\refcites}[1]{Refs.~\cite{#1}}
\newcommand\bets{\begin{table*}}
\newcommand\eets[1]{\label{tb:#1}\end{table*}}

\begin{document}
\preprint{LA-UR-24-29020}
\title{Moments of Axial-Vector GPD from Lattice QCD: \\ 
Quark Helicity, Orbital Angular Momentum, and Spin-Orbit Correlation}

\author[a]{Shohini Bhattacharya}
\author[b]{Krzysztof Cichy}
\author[c]{Martha Constantinou}
\author[d]{Xiang Gao}
\author[c]{Andreas Metz}
\author[c]{Joshua Miller}
\author[d]{Swagato Mukherjee}
\author[d]{Peter Petreczky}
\author[e]{Fernanda Steffens}
\author[f]{Yong Zhao}

\affiliation[a]{Theoretical Division, Los Alamos National Laboratory, Los Alamos, New Mexico 87545, USA}
\affiliation[b]{Faculty of Physics and Astronomy, Adam Mickiewicz University,
ul. Uniwersytetu Poznanskiego 2, 61-614 Poznan, Poland}
\affiliation[c]{Department of Physics,  Temple University,  Philadelphia,  PA 19122 - 1801,  USA}
\affiliation[d]{Physics Department, Brookhaven National Laboratory, Upton, New York 11973, USA}
\affiliation[e]{Institut f\"ur Strahlen- und Kernphysik, Rheinische Friedrich-Wilhelms-Universit\"at Bonn,\\ Nussallee 14-16, 53115 Bonn}
\affiliation[f]{Physics Division, Argonne National Laboratory, Lemont, IL 60439, USA}

\abstract{
In this work, we present a lattice QCD calculation of the Mellin moments of the twist-2 axial-vector generalized parton distribution (GPD), $\widetilde{H}(x,\xi,t)$, at zero skewness, $\xi$, with multiple values of the momentum transfer, $t$. Our analysis employs the short-distance factorization framework on ratio-scheme renormalized quasi-GPD matrix elements. The calculations are based on an $N_f=2+1+1$ twisted mass fermions ensemble with clover improvement, a lattice spacing of $a = 0.093$ fm, and a pion mass of $m_\pi = 260$ MeV. We consider both the iso-vector and iso-scalar cases, utilizing next-to-leading-order perturbative matching while omitting the disconnected contributions and gluon mixing in the iso-scalar case. For the first time, we determine the Mellin moments of $\widetilde{H}$ up to the fifth order. From these moments, we discuss the quark helicity and orbital angular momentum contributions to the nucleon spin, as well as the spin-orbit correlations of the quarks. Additionally, we perform a Fourier transform over the momentum transfer, which allows us to explore the spin structure in the impact-parameter space.
}

\maketitle
\flushbottom

\section{Introduction}

Understanding the internal structure of hadrons, such as protons and neutrons, is a fundamental goal in modern particle and nuclear physics. Generalized parton distributions (GPDs)~\cite{Muller:1994ses, Ji:1996ek, Radyushkin:1996nd} have emerged as a powerful tool for probing the three-dimensional structure of hadrons. Unlike traditional parton distribution functions (PDFs), which provide information about the longitudinal momentum distribution of partons, GPDs offer a more comprehensive picture by incorporating both longitudinal momentum and transverse spatial distributions~\cite{Burkardt:2000za, Ralston:2001xs, Diehl:2002he, Burkardt:2002hr}. Therefore, GPDs bridge the gap between the spatial and momentum distributions of quarks and gluons within the nucleon. 

The GPDs can be classified into different types based on their symmetry properties twist, and the polarization state of the parton/hadron~\cite{Diehl:2003ny,Ji:2004gf,Belitsky:2005qn,Boffi:2007yc,Mueller:2014hsa, Kumericki:2016ehc}. Among these, the twist-2 axial vector GPD $\widetilde{H}(x,\xi,t)$ is crucial for the understanding of the nucleon spin structure, a major challenge in hadronic physics~\cite{EuropeanMuon:1987isl,EuropeanMuon:1989yki}. The first Mellin moment of this GPD is directly related to the quark helicity contribution to the nucleon spin and, through Ji's spin decomposition scheme~\cite{Ji:1996ek}, provides insights into the orbital angular momentum (OAM). Additionally, it sheds light on spin-orbit correlations of quarks~\cite{Lorce:2014mxa,Rajan:2017cpx,Rodekamp:2023wpe}, offering valuable perspectives on the spin structure of the nucleon.

In principle, information on the GPDs can be obtained from experimental data for hard exclusive scattering processes such as the deeply virtual Compton scattering~\cite{Mueller:1998fv, Ji:1996ek, Radyushkin:1996nd, Ji:1996nm, Collins:1998be}, deeply virtual meson production~\cite{Radyushkin:1996ru, Collins:1996fb, Mankiewicz:1997uy} and processes where additional particles are detected in the final state~\cite{Pedrak:2017cpp,Duplancic:2018bum, Qiu:2022bpq, Qiu:2022pla, Duplancic:2023kwe,Qiu:2023mrm}. However, extracting GPDs from such data are highly non-trivial as it requires solving an inverse problem and disentangling multi-dimensional distributions from limited experimental observables~\cite{Bertone:2021yyz, Moffat:2023svr}. While much progress has been made in recent years~\cite{Cuic:2020iwt,Kriesten:2021sqc,Hashamipour:2021kes,Guo:2022upw,Guo:2023ahv,Hashamipour:2022noy, Irani:2023lol,Cuic:2023mki},
the field is still in its infancy. Consequently, computing GPDs from first principles using lattice QCD is well motivated, as it provides essential complementary information for constraining the GPDs.

Unfortunately, the direct simulation of GPDs is forbidden on a Euclidean lattice as they are defined through non-local light-cone correlators. Therefore, for a long time the focus has been on the Mellin moments of GPDs which can be computed through the matrix elements of local operators~\cite{Hagler:2003jd, QCDSF-UKQCD:2007gdl, Alexandrou:2011nr, Alexandrou:2013joa,Constantinou:2014tga,Green:2014xba,Alexandrou:2017ypw,Alexandrou:2017hac,Hasan:2017wwt,Gupta:2017dwj,Capitani:2017qpc,Alexandrou:2018sjm,Shintani:2018ozy,Bali:2018qus,Bali:2018zgl,Alexandrou:2019ali,Jang:2019jkn,Constantinou:2020hdm,Alexandrou:2022dtc,Jang:2023zts}. However, this method encounters difficulties in accessing higher moments due to signal decay and operator mixing under renormalization, which may be mitigated through the application of gradient flow~\cite{Shindler:2023xpd,Monahan:2016bvm} or smearing~\cite{Davoudi:2012ya}. Over the past few years, significant progress has been made in computing parton distributions using alternative methods~\cite{Liu:1993cv,Aglietti:1998ur,Detmold:2005gg,Braun:2007wv,Ji:2013dva,Ji:2014gla,Chambers:2017dov,Radyushkin:2017cyf, Orginos:2017kos,Ma:2014jla,Detmold:2021uru,Gao:2023lny}, especially those motivated by the proposal of quasi-PDFs~\cite{Ji:2013dva,Ji:2014gla}. Starting from boosted non-local equal-time correlators, the $x$-dependent parton distributions and their moments can be extracted using the framework of large momentum effective theory (LaMET)~\cite{Ji:2020ect} or short distance factorization (SDF)~\cite{Braun:2007wv,Radyushkin:2017cyf,Orginos:2017kos,Ma:2014jla}. For reviews, see \refcites{Cichy:2018mum,Ji:2020ect,Constantinou:2020pek,Cichy:2021lih,Cichy:2021ewm, Gao:2024pia}.

There has been a lot of  progress in computing GPDs through LaMET and SDF in the past few years~\cite{Karpie:2018zaz,Chen:2019lcm,Alexandrou:2020zbe,Lin:2020rxa,Alexandrou:2021bbo,CSSMQCDSFUKQCD:2021lkf,Bhattacharya:2021oyr,Lin:2021brq,Bhattacharya:2022aob,Constantinou:2022fqt,Bhattacharya:2023tik,Cichy:2023dgk,Bhattacharya:2023ays,Bhattacharya:2023nmv,Bhattacharya:2024qpp,HadStruc:2024rix,Ding:2024umu,Ding:2024hkz}. However, establishing these three-dimensional distributions with comprehensive $x$, $\xi$, and $t$ dependence remains challenging due to the computational cost. Significant progress was made recently in reducing these costs, as originally proposed in \refcite{Bhattacharya:2022aob}. By employing the Lorentz-covariant parametrization of matrix elements, quasi-GPDs can be constructed from Lorentz-invariant amplitudes determined from any reference frame. In particular, this innovative approach allows calculations in an asymmetric frame, applying all momentum transfer to the initial-state or final-state nucleon, rather than the commonly used symmetric frame. Consequently, multiple momentum transfers can be achieved through contractions without the need for additional inversions, leading to a faster and more efficient computation of GPDs using lattice QCD. In \refcites{Cichy:2023dgk,Bhattacharya:2023ays}, we presented the $x$-dependent twist-2 GPDs for unpolarized quarks, specifically $H$ and $E$, across multiple values of the momentum transfer $t$, along with their moments up to the fifth order. In \refcite{Bhattacharya:2023jsc}, we extend the theoretical framework to the case of axial-vector GPDs. Building on this progress, in the present work, we extract the Mellin moments of the zero-skewness axial-vector GPD $\widetilde{H}(x,0,t)$ over a wide range of $t$, and we discuss the physical insights that these moments provide.

This work is organized as follows. In \sec{framework}, we review the theoretical framework of computing the quasi axial-vector GPD on the lattice. In \sec{mx}, we show the bare matrix elements of the axial-vector iso-vector and iso-scalar GPDs and discuss the renormalization. In \sec{SDF}, we extract the first few moments from the ratio-scheme renormalized matrix elements using the next-to-leading order (NLO) SDF formula. For the first time, we get access up to the fifth moment of the axial-vector GPD $\widetilde{H}(x,\xi,t)$ with reasonable signal and $t$ dependence. In \sec{spin}, we discuss the relation between the moments and the spin structure of the nucleon, including the quark helicity and OAM contributions to the nucleon spin as well as the quark spin-orbit correlations. We also explore the distribution of these quantities in the impact parameter plane. Finally, \sec{conclusion} contains our conclusions.

\section{Axial-vector GPD on the lattice}\label{sec:framework}

\subsection{The definition of axial-vector GPDs}
The quark GPDs of nucleon are defined as the Fourier transform of the off-forward matrix elements
\begin{align}\label{eq:GPDdef}
F^{[\Gamma]}(z^-, \Delta, P)=\langle p_f;\lambda'| O_\Gamma|p_i;\lambda \rangle,    
\end{align}
where $p_i$ and $p_f$ represent the momenta of the initial-state and final-state nucleon, respectively, while $\lambda$ and $\lambda'$ denote the helicities of the nucleons. After performing the Fourier transform over $z^-$, the GPDs become functions of the average longitudinal momentum fraction $x$ of the quarks and two additional kinematic variables, typically chosen as the skewness $\xi$ and the momentum transfer squared $t$. Using $P = (p_i + p_f)/2$ and $\Delta = p_f - p_i$, they are defined as
\begin{align}
\xi=-\frac{\Delta^+}{2P^+},\quad  t=\Delta^2.
\end{align}
The quark bilinear operator involved is defined as,
\begin{align}
    O_\Gamma= \bar{\psi} (-\tfrac{z^-}{2}) \Gamma  {\cal W}(-\tfrac{z^-}{2},\tfrac{z^-}{2}) \psi (\tfrac{z^-}{2}),
\end{align}
where the quark fields are separated along the light-cone and connected by a Wilson line to ensure gauge invariance,
\begin{align}
 {\cal W}(-\tfrac{z^-}{2},\tfrac{z^-}{2}) = {\cal P} \, {\rm exp} \bigg ( -ig \int^{\tfrac{z^{-}}{2}}_{-\tfrac{z^{-}}{2}} \, dy^{-} A^{+}(0^{+},y^{-},\vec{0}_{\perp}) \bigg ) \,. 
\label{eq:wilson_line_standard_GPD}
\end{align}
In the light-cone gauge $A^+=0$, the Wilson line vanishes, allowing the operator $O_\Gamma$ to be interpreted as a particle density operator. For example, setting $\Gamma=\gamma^+(1+\gamma_5)/2$ and $\gamma^+(1-\gamma_5)/2$ corresponds to the density of right-handed and left-handed quarks inside the hadron, respectively. When summed, they give $\Gamma=\gamma^+$, providing the total quark number density. Their difference ($\Gamma=\gamma^+\gamma_5$) yields the quark helicity density, a crucial quantity for understanding the spin structure of the hadron.

In this work, we study the axial-vector GPDs defined through $\Gamma=\gamma^+\gamma_5$. At the twist-2 level, there are two distinct axial-vector GPDs, $\widetilde{H}$ and $\widetilde{E}$, defined through~\cite{Diehl:2003ny}
\begin{align}
F^{[\gamma^+ \gamma_5]} (z^-, \Delta, P)&  =  \bar{u}(p_f, \lambda ') \bigg [ \gamma^+ \gamma_5 \widetilde{H} (z^-, \xi, t) +  \frac{\Delta^+ \gamma_5}{2m} \widetilde{E} (z^-, \xi, t) \bigg ] u(p_i, \lambda) \, .
\label{eq:GPD_def_pos}
\end{align}
Since we are focusing on $\xi = 0$ case, the kinematic factor associated with $\widetilde{E}$ vanishes, implying that this GPD cannot be extracted from our lattice data. Here, we therefore concentrate on the GPD $\widetilde{H}$. We note that $\widetilde{E}$ also enters at the twist-3 level, where it can be addressed even for $\xi = 0$~\cite{Kiptily:2002nx, Bhattacharya:2023nmv}.

\subsection{The axial-vector quasi GPD $\widetilde{\cal H}$}\label{sec:quasiH}

The light-cone GPDs can be accessed from lattice QCD through the quasi-GPD approach in the large momentum limit. However, unlike light-cone GPDs, which are frame-independent, quasi-GPDs are frame-dependent at finite momentum. Traditionally, a specific symmetric frame was chosen.
However, this choice is computationally very expensive, requiring separate, full computations—including inversions and contractions on the lattice—for each value of $t$. To address this problem, we proposed constructing quasi-GPDs using Lorentz-invariant amplitudes derived from the decomposition of the matrix elements~\cite{Bhattacharya:2022aob}. This method eliminates the frame dependence, allowing any computationally preferred frame to be used. It is important to note that this decomposition is not unique and depends on the choice of basis. However, any basis will result in the same number of independent amplitudes. For the axial current we adopted the following decomposition~\cite{Bhattacharya:2023jsc},
\begin{align}
{F}^{[\gamma^\mu \gamma_5]} (z, P, \Delta)
& \equiv \langle p_f;\lambda'| \bar{\psi} (-\tfrac{z}{2})\, \gamma^\mu \gamma_5 \, {\cal W}(-\tfrac{z}{2},\tfrac{z}{2}) \psi (\tfrac{z}{2})|p_i;\lambda \rangle \nonumber \\[0.3cm]
& = \bar{u}(p_f,\lambda') \bigg [ \dfrac{i \epsilon^{\mu P z \Delta}}{m} \widetilde{A}_1 + \gamma^{\mu} \gamma_5 \widetilde{A}_2 + \gamma_5 \bigg ( \dfrac{P^\mu}{m} \widetilde{A}_3 + m z^\mu \widetilde{A}_4 + \dfrac{\Delta^\mu}{m} \widetilde{A}_5 \bigg ) \nonumber \\[0.1cm]
& \hspace{1.65cm} + m \slashed{z}\gamma_5 \bigg ( \dfrac{P^\mu}{m} \widetilde{A}_6 + m z^\mu \widetilde{A}_7 + \dfrac{\Delta^\mu}{m} \widetilde{A}_8 \bigg )\bigg ] u(p_i, \lambda),
\label{eq:helicity_para}
\end{align}
where $\epsilon^{\mu P z \Delta}=\epsilon^{\mu \alpha \beta \gamma} P_\alpha z_\beta \Delta_\gamma$ and $\widetilde{A}_i \equiv \widetilde{A}_i (z\cdot P, z \cdot \Delta, \Delta^2, z^2)$ are Lorentz invariant amplitudes that depend on Lorentz scalars. The case of the light-cone axial-vector GPDs defined in \Eq{GPD_def_pos} corresponds to $\mu=+$ and $z=(z^+,z^-,z_\perp)=(0,z^-,0_\perp)$.  Thus, the GPD $\widetilde{H}$ can be expressed as
\begin{align}
\widetilde{H}(z\cdot P, z \cdot \Delta, \Delta^2)  & = \widetilde{A}_2 + (P^+ z^-) \widetilde{A}_6 + (\Delta^+ z^-) \widetilde{A}_8 \nonumber\\
& = \widetilde{A}_2 + (P \cdot z) \widetilde{A}_6 + (\Delta \cdot z) \widetilde{A}_8.\label{eq:H_LC}
\end{align}

As mentioned earlier, light-cone GPDs cannot be directly simulated in Euclidean lattice QCD. In this work, we consider quasi-GPDs, which maintain the same form as \Eq{GPD_def_pos} but are defined at equal time ($z^0=0$). In this approach, quarks are separated along the spatial direction $\mathbf{z}=(0,0,z^3)$ with a large momentum $\mathbf{P}=(0,0,P^3)$. Typically, $\gamma^\mu\gamma_5=\gamma^3\gamma_5$ is chosen to approach the light-cone limit, as it avoids operator mixing caused by explicit chiral symmetry breaking which affects $\gamma^0\gamma_5$~\cite{Constantinou:2017sej,Chen:2017mzz}.
This means, we consider
\begin{align}
F^{[\gamma^3 \gamma_5]} (z, P,\Delta)&  =  \bar{u}(p_f, \lambda ') \bigg [ \gamma^3 \gamma_5 \widetilde{\cal H}_3 (z, \xi, t) +  \frac{\Delta^3 \gamma_5}{2m} \widetilde{\cal E}_3 (z, \xi, t) \bigg ] u(p_i, \lambda) \, .
\label{eq:qusi_GPD_def}
\end{align}
According to \Eq{helicity_para}, the axial-vector quasi GPDs can be expressed as
\begin{align}
\label{eq:qH3}
\widetilde{\mathcal{H}}_3 (z, P, \Delta)  & = \widetilde{A}_2 - z^3 P^3 \widetilde{A}_6 - m^2 (z^3)^2 \widetilde{A}_7 - z^3 \Delta^3 \widetilde{A}_8
\nonumber \\
& =\widetilde{A}_2 + (P \cdot z) \widetilde{A}_6 + m^2 z^2 \widetilde{A}_7 + (\Delta \cdot z) \widetilde{A}_8.
\end{align}
Compared to \Eq{H_LC}, the Lorentz invariant amplitudes $\widetilde{A}_i (z\cdot P, z \cdot \Delta, \Delta^2, z^2)$ in $\widetilde{\mathcal{H}}_3$ implicitly depend on the finite $z^2=-|\mathbf{z}|^2$, which is zero in the light-cone case. In addition, $\widetilde{\mathcal{H}}_3$ explicitly includes an additional contamination term $m^2 z^2 \widetilde{A}_7$ due to the non-vanishing $z^2$. In \refcites{Bhattacharya:2022aob, Bhattacharya:2023jsc}, it was proposed to remove these explicit power corrections and construct Lorentz-invariant (LI) matrix elements, as in \Eq{H_LC}, through the Lorentz-invariant amplitudes $\widetilde{A}_i$, which can be extracted from the linear combination of quasi-GPDs with different spin structures. As discussed in detail in \refcite{Bhattacharya:2023jsc}, the LI axial-vector GPDs in \Eq{H_LC} can be derived from the linear combination of $F^{[\gamma^\mu \gamma_5]}$ with $\mu=0,1,2$, referred to as $\widetilde{\mathcal{H}}$. We note that the only difference between the quasi GPD $\widetilde{\mathcal{H}}$ and light-cone GPD $\widetilde{{H}}$ is the non-zero $z^2$ in $\widetilde{A}_i$. Interestingly, it was found in \refcite{Bhattacharya:2023jsc} that the term $m^2 z^2 \widetilde{A}_7$ is mostly consistent with zero, so that $\widetilde{\mathcal{H}}_3$ and 
$\widetilde{\mathcal{H}}$ are largely consistent within statistical errors. Therefore, we can focus on $\widetilde{\mathcal{H}}_3$ in the following analysis without concerns about the contamination term $m^2 z^2 \widetilde{A}_7$. Additionally, we repeat that $\widetilde{\mathcal{H}}_3$ has the advantage of avoiding operator mixing due to explicit chiral symmetry breaking from lattice discretization~\cite{Constantinou:2017sej,Chen:2017mzz}.

\section{Bare matrix elements and renormalization}\label{sec:mx}

\subsection{Lattice setup}
The data used in this work has been analyzed in \refcite{Bhattacharya:2023jsc}, to derive the $x$-dependent GPD in the LaMET framework. In this study, however, we focus on extracting the first few moments of GPDs using the SDF approach. The data were obtained from a gauge ensemble of $N_f=2+1+1$ twisted-mass fermions with a clover term and Iwasaki-improved gluons~\cite{Alexandrou:2018egz}. The lattice size and spacing of the ensemble are $N_s\times N_t=32^3 \times 64$ and $a$ = 0.0934 fm, respectively, with quark masses corresponding to a pion mass of 260 MeV.

\begin{table}[h!]
\begin{center}
\renewcommand{\arraystretch}{1.9}
\begin{tabular}{lcccc|cccc}
\hline
frame & $P_3$ [GeV] & $\quad \mathbf{\Delta}$ $[\frac{2\pi}{L}]\quad$ & $-t$ [GeV$^2$] & $\quad \xi \quad $ & $N_{\rm ME}$ & $N_{\rm confs}$ & $N_{\rm src}$ & $N_{\rm tot}$\\
\hline
N/A       & $\pm$1.25 &(0,0,0)  &0   &0   &2   &329  &16  &10528 \\
\hline
symm      & $\pm$0.83 &($\pm$2,0,0), (0,$\pm$2,0)  &0.69   &0   &8   &67 &8  &4288 \\
symm      & $\pm$1.25 &($\pm$2,0,0), (0,$\pm$2,0)  &0.69   &0   &8   &249 &8  &15936 \\
symm      & $\pm$1.67 &($\pm$2,0,0), (0,$\pm$2,0)  &0.69   &0   &8   &294 &32  &75264 \\
symm      & $\pm$1.25 &$(\pm 2,\pm 2,0)$           &1.38   &0   &16   &224 &8  &28672 \\
symm      & $\pm$1.25 &($\pm$4,0,0), (0,$\pm$4,0)  &2.77   &0   &8   &329 &32  &84224 \\
\hline
asymm  & $\pm$1.25 &($\pm$1,0,0), (0,$\pm$1,0)  &0.17   &0   &8   &269 &8  &17216\\
asymm      & $\pm$1.25 &$(\pm 1,\pm 1,0)$       &0.34   &0   &16   &195 &8  &24960 \\
asymm  & $\pm$1.25 &($\pm$2,0,0), (0,$\pm$2,0)  &0.65   &0   &8   &269 &8  &17216\\
asymm      & $\pm$1.25 &($\pm$1,$\pm$2,0), ($\pm$2,$\pm$1,0) &0.81   &0   &16   &195 &8  &24960 \\
asymm  & $\pm$1.25 &($\pm$2,$\pm$2,0)          &1.24    &0   &16  &195 &8   &24960\\
asymm  & $\pm$1.25 &($\pm$3,0,0), (0,$\pm$3,0)  &1.38   &0   &8   &269 &8  &17216\\
asymm      & $\pm$1.25 &($\pm$1,$\pm$3,0), ($\pm$3,$\pm$1,0)  &1.52   &0   &16   &195 &8  &24960 \\
asymm  & $\pm$1.25 &($\pm$4,0,0), (0,$\pm$4,0)  &2.29   &0   &8   &269 &8  &17216\\
\hline
\end{tabular}
\caption{\small Statistics for the symmetric and asymmetric frame matrix elements are shown. The momentum unit $2\pi/L$ is 0.417 GeV. $N_{\rm ME}$, $N_{\rm confs}$, $N_{\rm src}$ and $N_{\rm total}$ are the number of matrix elements, configurations, source positions per configuration and total statistics, respectively.}
\label{tb:stat}
\end{center}
\end{table}

The quasi-GPD matrix elements are extracted from the three-point functions,
\begin{align}
\begin{split}
    C^{\rm 3pt}_\mu (\Gamma_\kappa, p_f, p_i; t_s, \tau) = \sum_{\yvec, \zvec_0} e^{-i \pvec_f \cdot (\yvec - \xvec)} e^{-i \qvec \cdot (\xvec - \zvec_0)} \Gamma^\kappa_{\alpha \beta}\langle{N_\alpha^{(s)} (\yvec, t_s) \mathcal{O}_\mu (\zvec_0 + z \hat{z}, \tau) \overline{N}_\beta^{(s)} (\xvec, 0)} \rangle,
\end{split}
\end{align}
where $\vec x$ is the source position, $N^{(s)}$ is the standard nucleon source under momentum smearing~\cite{Bali:2016lva} to improve the overlap with the proton ground state and suppress gauge noise. The quasi-GPD operator $\mathcal{O}_\mu=\bar\psi\left(z\right) \gamma^\mu\gamma_5 {\cal W}(z,0)\psi\left(0\right)$ has quark fields separated along the $z_3$ direction. Both the iso-vector ($u-d$) and iso-scalar ($u+d$) flavor combinations were computed with the disconnected diagrams ignored for the iso-scalar case. In \refcite{Alexandrou:2021oih}, it was found that, on the same ensemble as this work, the disconnected contributions for the forward limit are tiny; they would be further suppressed in off-forward kinematics. The unpolarized and polarized parity projectors $\Gamma_0$ and $\Gamma_\kappa$ are defined as,
\begin{align}
	\Gamma_0 &= \frac{1}{4} \left(1 + \gamma_0\right)\,, \\
	\Gamma_\kappa &= \frac{1}{4} \left(1 + \gamma_0\right) i \gamma_5 \gamma_\kappa\,, \quad \kappa=1,2,3\,.
\end{align}
To derive the ground-state matrix elements, we also computed the two-point functions for the energy spectrum and overlap amplitudes $\langle \Omega|N^{(s)}|N\rangle$, which are given by,
\begin{align}
\begin{split}
    C^{\rm 2pt}(\Gamma_0, p;t_s) = \sum_{\yvec} e^{-i \pvec \cdot (\yvec - \xvec)} \Gamma^0_{\alpha \beta} \langle {N^{(s)}_\alpha (\yvec, t_s ) \overline{N}^{(s^\prime)}_\beta (\xvec, 0)} \rangle\,.
\end{split}
\end{align}
Since the two- and three-point functions are highly correlated, we construct the ratio,
\begin{align}
R^\mu_\kappa (\Gamma_\kappa, p_f, p_i; t_s, \tau) = \frac{C^{\rm 3pt}_\mu (\Gamma_\kappa, p_f, p_i; t_s, \tau)}{C^{\rm 2pt}(\Gamma_0, p_f;t_s)} \sqrt{\frac{C^{\rm 2pt}(\Gamma_0, p_i, t_s-\tau)C^{\rm 2pt}(\Gamma_0, p_f, \tau)C^{\rm 2pt}(\Gamma_0, p_f, t_s)}{C^{\rm 2pt}(\Gamma_0, p_f, t_s-\tau)C^{\rm 2pt}(\Gamma_0, p_i, \tau)C^{\rm 2pt}(\Gamma_0, p_i, t_s)}}\,,
\end{align}
which, in the $t_s\rightarrow\infty$ limit, corresponds to the bare matrix elements of proton ground state $\lim\limits_{t_s\rightarrow\infty} R^\mu_\kappa=\Pi_\mu(\Gamma_\kappa)$. To keep the statistical noise under control, we use a source-sink separation of $t_s = 10a = 0.93$ fm and perform a plateau fit with respect to the time insertion $\tau$ in a region of convergence.  More details can be found in \refcite{Bhattacharya:2023jsc}. A more thorough study of excited state contamination will be left for future work that targets precision control.

In \tb{stat}, we show the momenta $\vec{P}=(0,0,P_3)$ and $\vec{\Delta}$ as well as the statistics used in this work. For the symmetric frame, the momentum components are defined as,
\begin{equation}
\label{eq:pf_symm}
\vec{p}^{\,s}_f=\vec{P} + \frac{\vec{\Delta}}{2} = \left(+\frac{\Delta_1}{2},+\frac{\Delta_2}{2},P_3\right)\,,\qquad
\vec{p}^{\,s}_i=\vec{P} - \frac{\vec{\Delta}}{2}= \left(-\frac{\Delta_1}{2},-\frac{\Delta_2}{2},P_3\right)\,,
\end{equation}
In contrast, for the asymmetric frame, where all momentum transfer is assigned to the initial state, we have,
\begin{equation}
\vec{p}^{\,a}_f=\vec{P} =  \left(0,0,P_3\right) \,,\qquad
\label{eq:pi_nonsymm}
\vec{p}^{\,a}_i=\vec{P} - \vec{\Delta} =  \left(-\Delta_1,-\Delta_2,P_3\right)\,.
\end{equation}
While $\vec{P}$ and $\vec{\Delta}$ are the same for both frames, they lead to slightly different values of $-t$ due to the different distribution of the momentum transfer, that is
\begin{eqnarray}
-t^s = \vec{\Delta}^2\,,\qquad
-t^a = \vec{\Delta}^2 - (E(p')-E(p))^2    \,.
\label{eq:Qboosted}
\end{eqnarray}
This work focuses on zero skewness, namely $\Delta_3=0$. As already mentioned above, this does not give us access to the GPD $\widetilde{E}$. Most of the hadron momentum $P$ is fixed at 1.25 GeV throughout the calculation. We combine all data contributing to the same value of momentum transfer $t=-\Delta^2$ with definite symmetry with respect to $P_3 \to -P_3$, $z_3\to -z_3$, and $\vec{\Delta} \to -\vec{\Delta}$~\cite{Bhattacharya:2023jsc}.

\begin{figure}
    \centering
    \includegraphics[width=0.45\textwidth]{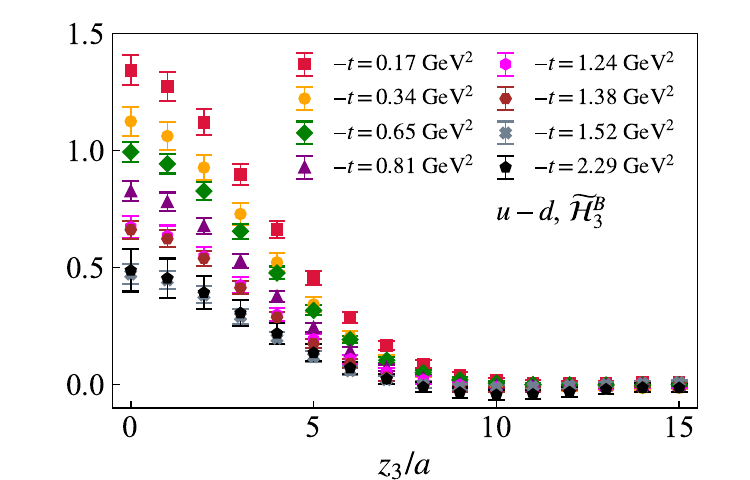}
    \includegraphics[width=0.45\textwidth]{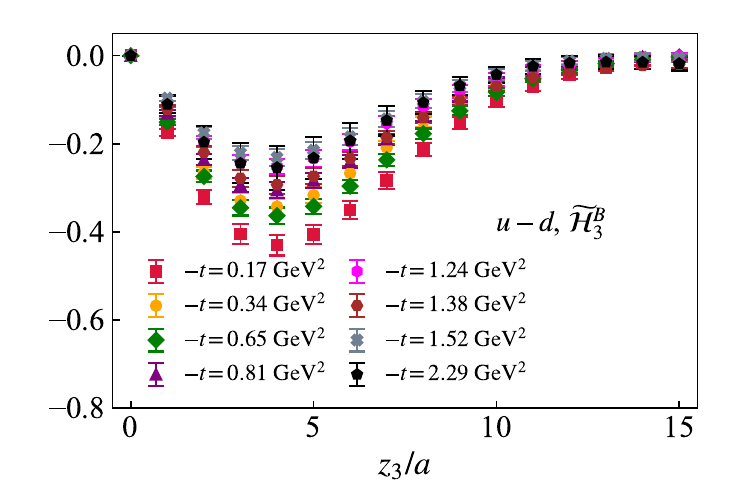}
    \includegraphics[width=0.45\textwidth]{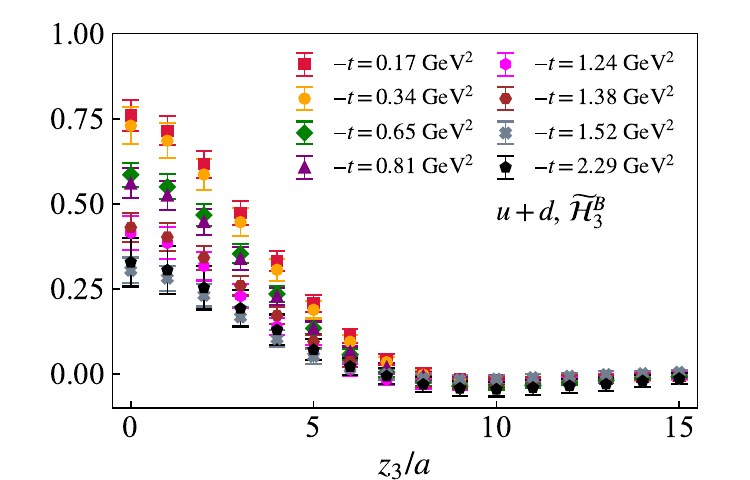}
    \includegraphics[width=0.45\textwidth]{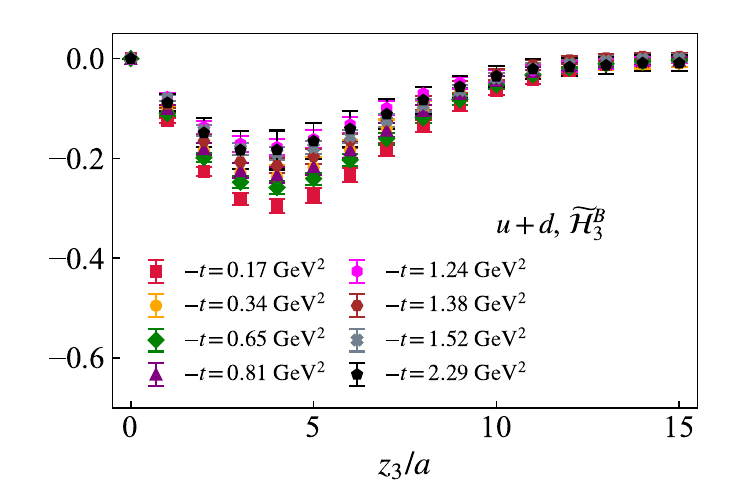}
	\caption{Bare matrix elements are shown as a function of $z_3/a$ for different values of $t$. The iso-vector ($u-d$) and iso-scalar ($u+d$) case are shown in the upper and lower panels, respectively. The left panels show the real part while the right panels show the imaginary part. \label{fig:H3}}
\end{figure}

\begin{figure}
    \centering
    \includegraphics[width=0.45\textwidth]{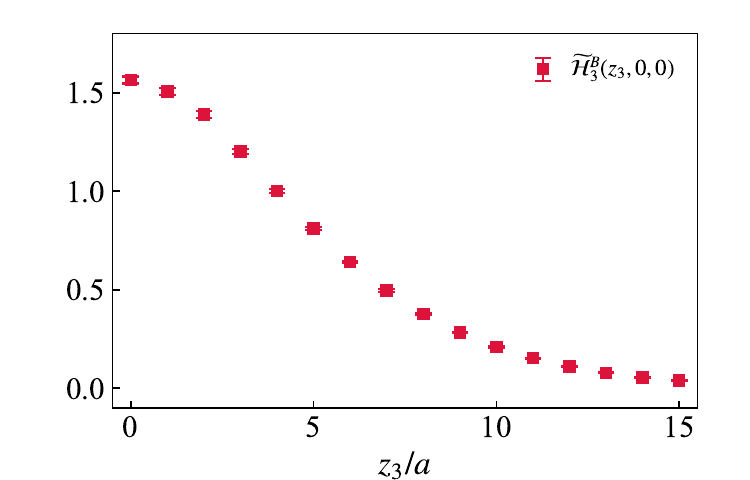}
	\caption{Zero momentum iso-vector axial-vector quasi-PDF matrix elements. \label{fig:Hp0}}
\end{figure}

\begin{figure}
    \centering
    \includegraphics[width=0.45\textwidth]{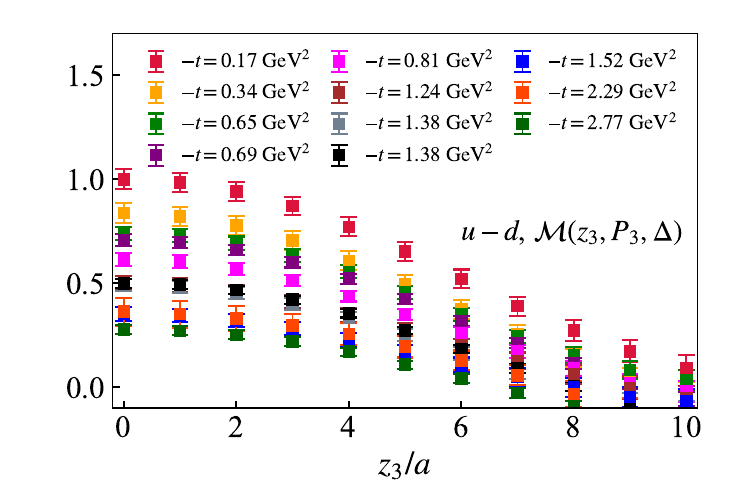}
    \includegraphics[width=0.45\textwidth]{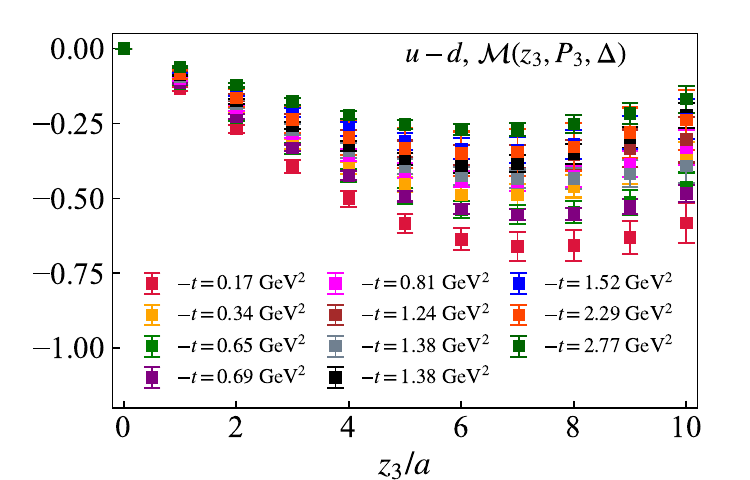}
    \includegraphics[width=0.45\textwidth]{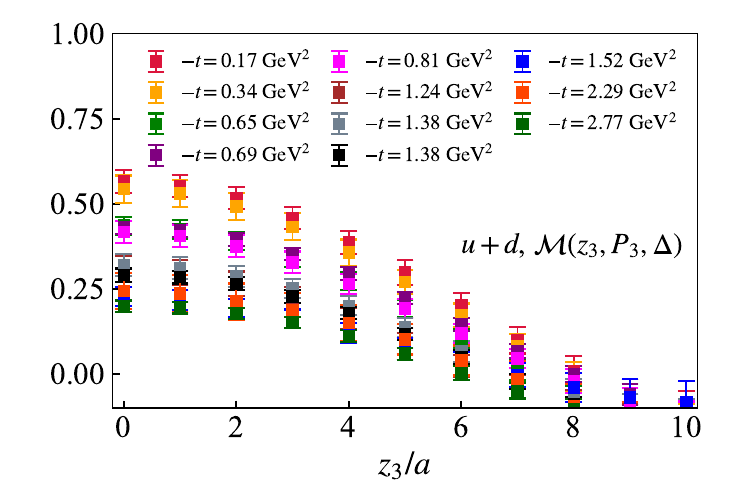}
    \includegraphics[width=0.45\textwidth]{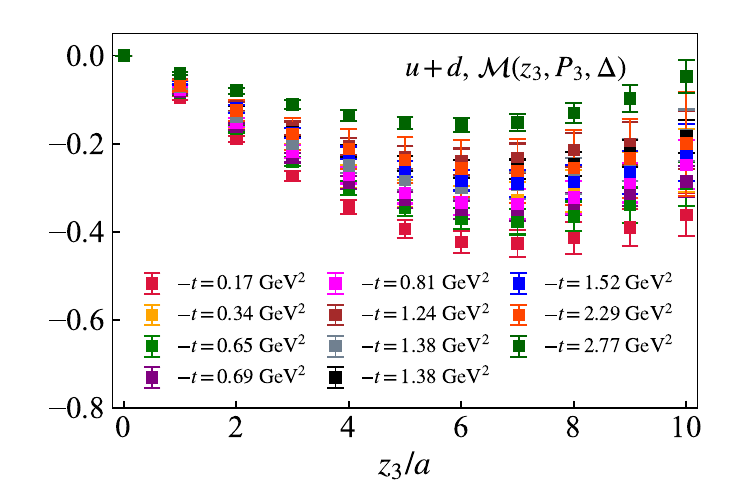}
	\caption{Ratio scheme renormalized matrix elements are shown as a function of $z_3/a$. The iso-vector ($u-d$) and iso-scalar ($u+d$) case are shown in the upper and lower panels, respectively. The left panels show the real part while the right panels show the imaginary part.\label{fig:rITDstd}}
\end{figure}

\subsection{Bare matrix elements and renormalization}

The bare matrix elements for the quasi axial-vector GPD $\widetilde{\mathcal{H}}_3$ are shown in \fig{H3}, where a clear signal is observed across a wide range of momentum transfers $-t$. The bare matrix elements need to be renormalized. At $z=0$ and $\Delta=0$, the iso-vector matrix element $\widetilde{\mathcal{H}}_3^B$ gives the bare iso-vector axial charge of the nucleon, $g_A^B$, which needs to be renormalized by the constant $Z_A$. The $Z_A$ for this ensemble has been determined as 0.7442~\cite{Bhattacharya:2023jsc}, leading to a derived $g_A=1.164(13)$ using $\widetilde{\mathcal{H}}_3^{B,u-d} (0, 0, 0)$. This value is consistent with results in \refcite{Alexandrou:2010hf} with similar quark masses. For the case of non-zero $z_3$, it has been shown that the non-local operator $O_\Gamma$ can be multiplicatively renormalized~\cite{Ji:2017oey,Green:2017xeu,Ishikawa:2017faj},
\begin{align}
    O_\Gamma(z) = Z_Oe^{-\delta m|z|}O_\Gamma^R(z),
\end{align}
where $Z_O$ accounts for logarithmic divergences, and the exponential factor $e^{-\delta m|z|}$ removes the linear divergence stemming from the self-energy of the spatial Wilson link. Since this renormalization is independent of the hadron state and quark flavor, one can construct an appropriate ratio to eliminate UV divergences and obtain renormalization group invariant quantities~\cite{Fan:2020nzz,Radyushkin:2017cyf, Orginos:2017kos,Gao:2020ito},
\begin{align}\label{eq:ratioH3}
    \mathcal{M}(z_3,P_3,\Delta)=\frac{\widetilde{\mathcal{H}}_3^B (z_3, P_3, \Delta;a)}{\widetilde{\mathcal{H}}_3^B (z_3, 0, 0;a)}\cdot g_A=\frac{\widetilde{\mathcal{H}}_3^R (z_3, P_3, \Delta;\mu)}{\widetilde{\mathcal{H}}_3^R (z_3, 0, 0;\mu)}\cdot g_A.
\end{align}
We multiply $g_A = Z_A \widetilde{\mathcal{H}}_3^{B,u-d}(0, 0, 0; a)$ to normalize the iso-vector $\mathcal{M}(0, 0, 0)$ to $g_A$. This choice of convention will be explained in the next section. In this ratio, the matrix elements in the denominator correspond to axial-vector quasi-PDF matrix elements with zero momentum $P=0$ and zero momentum transfer $\Delta=0$. Since UV divergences are independent of the light quark flavors, we consistently use the iso-vector $\widetilde{\mathcal{H}}_3^{B,u-d} (z_3, 0, 0)$ as the denominator in this work, which are shown in \fig{Hp0}. Both the iso-vector and iso-scalar $\widetilde{\mathcal{H}}_3^B (z, P, \Delta)$ are used in the numerator. It should be noted again that the disconnected diagrams were omitted for the iso-scalar case in this study.

In \fig{rITDstd}, the real and imaginary parts of the ratio scheme renormalized matrix elements are presented as a function of $z_3$ for both the iso-vector and iso-scalar cases. As one can see, the expected $-t$ dependence is clearly visible, with the magnitude of the matrix elements decreasing monotonically as $-t$ increases.

\section{Mellin moments from short distance factorization}\label{sec:SDF}

\subsection{Short distance factorization}
In the short-distance limit, the renormalized matrix elements can be expanded in terms of the Mellin moments using the operator product expansion (OPE). For the zero-skewness quasi-GPD under consideration, the OPE structure mirrors that of the quasi-PDF case, without any mixing between moments. In the $\overline{\rm MS}$ scheme, the short-distance factorization (SDF) of the iso-vector quasi-GPD matrix elements can be expressed as,
\begin{align}\label{eq:OPE}
\widetilde{\mathcal{H}}_3^R(z_3,P_3,\Delta;\mu)=\sum_{n=0}^{\infty}C_n^{\overline{\rm MS}}(\mu^2z^2)\frac{(-iz_3P_3)^n}{n!}\tilde{A}_{n+1,0}(t;\mu)+\mathcal{O}(\Lambda_{\rm QCD}^2z^2),
\end{align}
where the
\begin{align}
    \tilde{A}_{n+1,0}(t;\mu)=\int_{-1}^1dx x^n \widetilde{H}(x,\xi=0,t;\mu)
\end{align}
represent the Mellin moments of the axial-vector GPD at zero skewness. It's worth to mention that the first moment $\tilde{A}_{1,0}(t;\mu)$ is the nucleon axial form factor corresponding to the local matrix elements $\widetilde{\mathcal{H}}_3^R(0,P_3,\Delta;\mu)$. $C_n(\mu^2z^2)$ are the Wilson coefficients. At leading order (LO), $C_n(\mu^2z^2)=1$, making \Eq{OPE} a simple polynomial function of the so-called Ioffe time $\zeta=z_3P_3$. Beyond LO, the perturbative corrections account for the scale evolution from the physical scale $\sim 1/z_3$ to the factorization scale $\mu$, which at NLO are given by~\cite{Izubuchi:2018srq},
\begin{align}
\begin{split}
        C_n^{\overline{\rm MS}}(\mu^2z^2) = 1 + \frac{\alpha_s C_F}{2\pi}\bigg[\bigg(\frac{3+2n}{
2+3n+n^2}+2H_n\bigg)L_z+\frac{7+2n}{2+3n+n^2}+2(1-H_n)H_n-2H^{(2)}_n\bigg],
\end{split}
\end{align}
with $L_z=\ln(\mu^2z^2e^{2\gamma_E}/4)$ and the Harmonic numbers $H_n=\sum_{i=1}^n1/i$ and  $H_n^{(2)}=\sum_{i=1}^n1/i^2$. This SDF formula can be inserted into \Eq{ratioH3}, establishing a relationship between the renormalized matrix elements and the moments of the GPDs,
\begin{align}\label{eq:OPEratio}
\begin{split}
    \mathcal{M}(z_3,P_3,\Delta)&=\frac{\sum_{n=0}^{\infty}C_n^{\overline{\rm MS}}(\mu^2z^2)\frac{(-iz_3P_3)^n}{n!}\tilde{A}_{n+1,0}(t;\mu)+\mathcal{O}(\Lambda_{\rm QCD}^2z^2)}{C_0^{\overline{\rm MS}}(\mu^2z^2)\tilde{A}_{1,0}^{u-d}(0;\mu)+\mathcal{O}(\Lambda_{\rm QCD}^2z^2)}\cdot g_A\\
    &=\sum_{n=0}^{\infty}\frac{C_n^{\overline{\rm MS}}(\mu^2z^2)}{C_0^{\overline{\rm MS}}(\mu^2z^2)}\frac{(-iz_3P_3)^n}{n!}\tilde{A}_{n+1,0}(t)+\mathcal{O}(\Lambda_{\rm QCD}^2z^2).
\end{split}
\end{align}
We note again, the denominator in \Eq{ratioH3} is solely the iso-vector matrix element while the numerator can be either iso-vector and iso-scalar. As a result, $\tilde{A}_{1,0}^{u-d}(0;\mu)=g_A$ cancels out in the first line of the formula. This explains our choice of ratio defined in \Eq{ratioH3}. From this expression, the even and odd moments of the axial-vector GPD can be extracted from the real and imaginary parts of $\mathcal{M}(z_3,P_3,\Delta)$, respectively. Notably, it is crucial to keep $z^2$ small to avoid large power corrections. At present, lattice calculations are limited to finite values of $P_3$, which in our case are listed in \tb{stat}, allowing the extraction of only the first few moments within a limited kinematic range of $z_3P_3$.

\subsection{Moments from fixed $z^2$}

\begin{figure}
    \centering
    \includegraphics[width=0.45\textwidth]{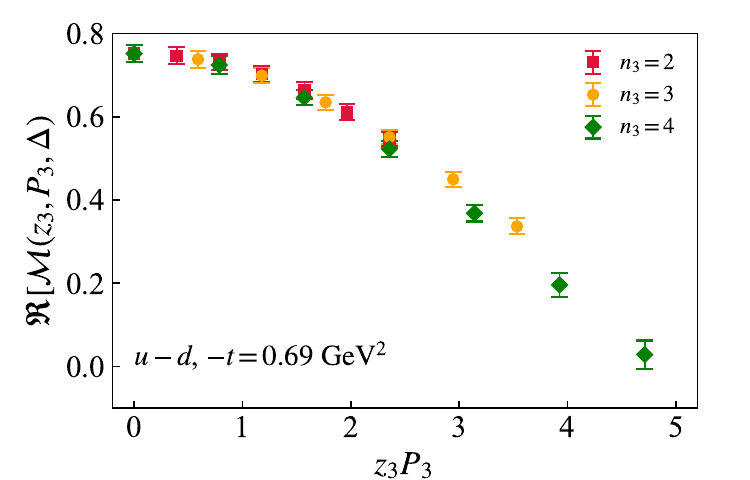}
    \includegraphics[width=0.45\textwidth]{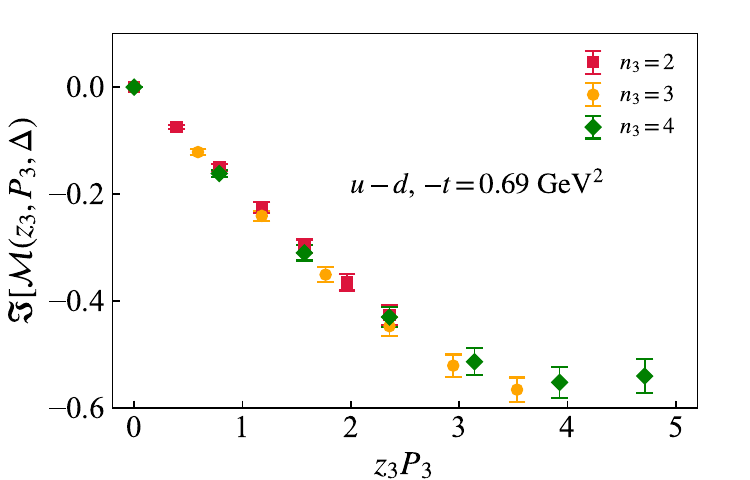}
	\caption{The ratio scheme renormalized iso-vector matrix elements for $-t=0.69~\rm GeV^2$ are shown as a function of $z_3P_3$ for three different values of $P_3$. The real part (left panel) and imaginary part (right panel) are both shown. \label{fig:rITDz3}}
\end{figure}

\begin{figure}
    \centering
    \includegraphics[width=0.45\textwidth]{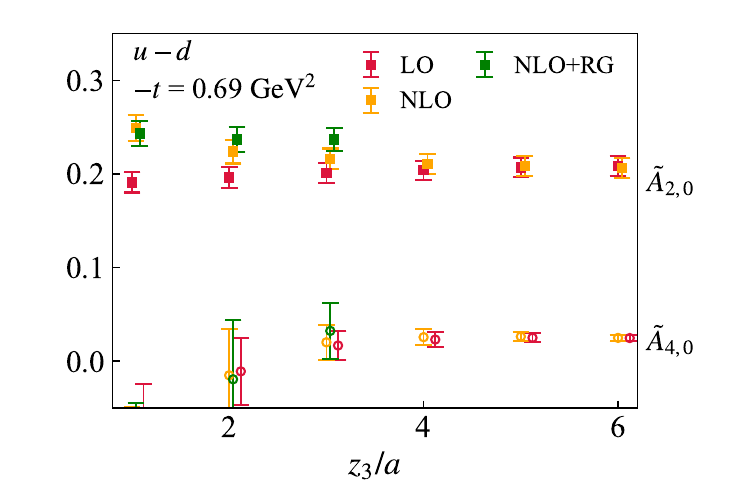}
    \includegraphics[width=0.45\textwidth]{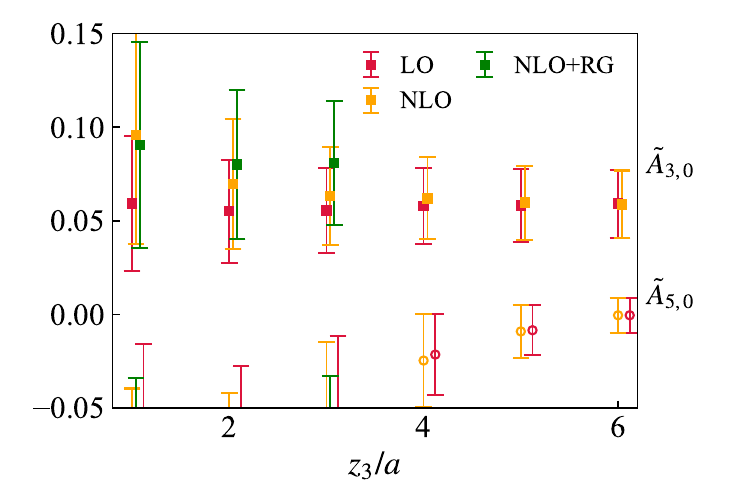}
	\caption{The iso-vector moments at momentum transfer $-t=0.69~\rm GeV^2$, derived from fits of $P_3$ dependence, are shown as a function of $z_3$.\label{fig:rITDz3moms}}
\end{figure}

For $-t=0.69\rm GeV^2$, we have three different momenta with $n_3 = 2, \, 3, \, 4$ corresponding to 0.83, 1.25, 1.67 GeV, respectively. 
This allows us to extract moments from each single $z_3$ by fitting the $P_3$ dependence. 
In this section, we consider the matching formula at LO, NLO, as well as 
renormalization group improved NLO (NLO+RG)~\cite{Gao:2021hxl,Su:2022fiu,Dutrieux:2023zpy,HadStruc:2024rix} accuracy. 
If the perturbative matching can describe the evolution well, the moments for a given factorization scale $\mu$ should be independent 
for different values of $z_3$.

In \fig{rITDz3}, we present the ratio scheme renormalized matrix elements for the iso-vector case as a function of $z_3 P_3$. One
can see from the plots that the dependence of the results on $P_3$ is very weak. This is expected because, when the perturbative evolution encoded in $C_n^{\overline{\rm MS}}(\mu^2z^2)$ and the power corrections are both small within the short $z_3$ range ($z_3\lesssim 3a$) and cancel to a good degree in the ratio, the renormalized matrix element should depend solely on $z_3 P_3$ within the current statistical error~\cite{Gao:2020ito,Bhattacharya:2023ays}. To extract the moments according to \Eq{OPEratio}, we minimize,
\begin{align}
    \chi^2_{z_3} &=\sum_{P_3}\bigg(\frac{(\Re[\mathcal{M}^{\rm data}(z_3,P_3,\Delta)]-\Re[\mathcal{M}^{\rm SDF}(z_3,P_3,\Delta)])^2}{\sigma_{\Re}^2} \nonumber\\
    & \qquad + \frac{(\Im[\mathcal{M}^{\rm data}(z_3,P_3,\Delta)]-\Im[\mathcal{M}^{\rm SDF}(z_3,P_3,\Delta)])^2}{\sigma_{\Im}^2}\bigg),
\end{align}
where $\sigma_{\Re}$ and $\sigma_{\Im}$ represent the errors in the real and imaginary parts of $\mathcal{M}^{\rm data}(z_3,P_3,\Delta)$, respectively.

The moments obtained from the fits are shown in \fig{rITDz3moms} as a function of $z_3$. The results are evaluated at $\mu=2$ GeV. We omit the discussion of the axial form factor $\tilde{A}_{1,0}$ here, as it is determined from local matrix elements, requiring neither the SDF nor dependence on $z_3$. For the first two non-trivial moments, $\tilde{A}_{2,0}$ and $\tilde{A}_{3,0}$, a reasonable signal emerges starting from $z_3=a$. However, for the higher moments a clear signal can only
be obtained for larger values of $z_3$. As one can see, $\tilde{A}_{2,0}$ exhibits a mild $z_3$ dependence for $z_3<4 a$ when LO and NLO Wilson coefficients are used. This mild dependence is due to a combination of discretization effects, which could be especially large for $z_3=a$,
and missing higher-order terms in the Wilson coefficients.
The results that use NLO Wilson coefficients with RG improved coefficients are only shown for $z_3$ up to $3a$, as for larger $z_3$ values the scale is too low to evaluate the strong coupling constant $\alpha_s$. In that range, the NLO results and NLO results with RG resummation are consistent.

\subsection{Moments from combined fits}\label{sec:moments}

\begin{figure}
    \centering
    \includegraphics[width=0.45\textwidth]{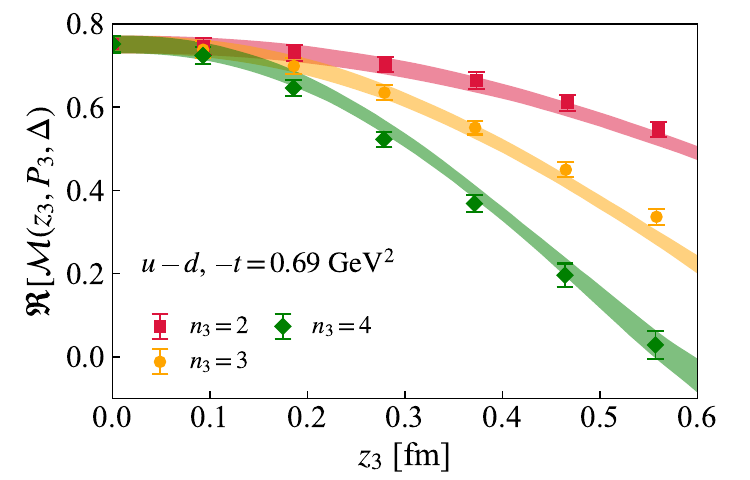}
    \includegraphics[width=0.45\textwidth]{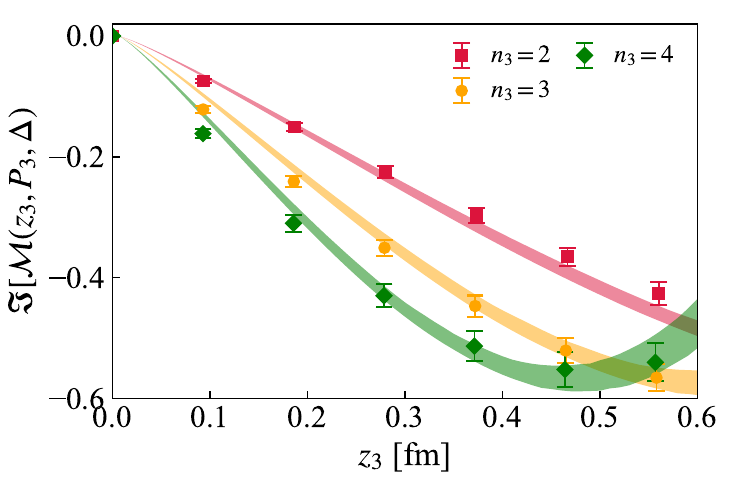}
	\caption{The ratio scheme renormalized iso-vector matrix elements for the case with momentum transfer $-t=0.69\rm GeV^2$ are shown as a function of $z_3$ for three different values of $P_3$. The real part (left panel) and imaginary part (right panel) are both shown. The bands are reconstructed from the fit with $z_3\in[2a,6a]$ including the NLO matching kernels. \label{fig:rITDfit}}
\end{figure}

\begin{figure}
    \centering
    \includegraphics[width=0.35\textwidth]{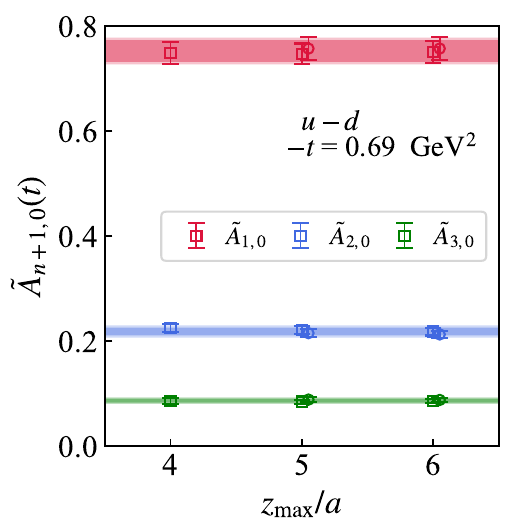}
    \includegraphics[width=0.35\textwidth]{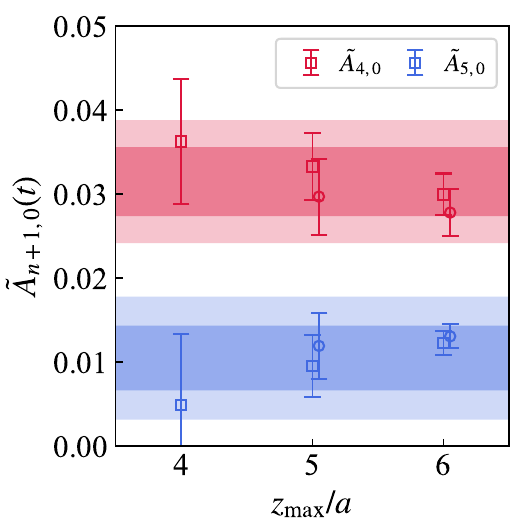}
	\caption{The moments obtained from combined fit using matrix element with $z_3\in[z_{\rm min}, z_{\rm max}]$ and $n_3=2,3,4$ are shown as a function of $z_{\rm max}$. The squared symbols are results from $z_{\rm min}=2a$ while the circled symbols are from $z_{\rm min}=3a$.\label{fig:momsCombz}}
\end{figure}

Since the factorization formula with NLO coefficient can describe the lattice data reasonably well, in this section we perform combined fits of matrix elements with different $n_3$ in a range of $z_3\in[z_{\rm min}, z_{\rm max}]$.  
Specifically, we minimize the $\chi^2=\sum_{z_3}\chi^2_{z_3}$. 
The combined fits are more stable than the fits for fixed $z_3$ values.
To avoid the most serious discretization effects, we skip the point $z_3=a$. We vary $z_{\rm min}\in[2a,3a]$ to estimate the discretization effects and $z_{\rm max}\in[z_{\rm min}+2a,6a]$ to estimate the higher-twist effects. The resulting fit yields a reasonable $\chi^2/{\rm d.o.f.}$, and the bands reconstructed from the fit accurately describe the renormalized matrix elements, as shown in \fig{rITDfit}.

The extracted moments are shown in \fig{momsCombz}. The squared symbols represent results from $z_{\rm min}=2a$, while the circled symbols correspond to $z_{\rm min}=3a$. As illustrated, the results from the two $z_{\rm min}$ values overlap, indicating that discretization effects are minimal compared to the statistical errors. Regarding the $z_{\rm max}$ dependence, it appears negligible within the errors, particularly for the lower moments. However, higher moments require larger $z_{\rm max}$ values to stabilize the fit.

For the final estimates, we average the results obtained from different choices of $z_{\rm min}$ and $z_{\rm max}$, with their deviations treated as systematic errors, as we did in \refcites{Gao:2020ito,Bhattacharya:2023ays}. These results are depicted as bands in \fig{momsCombz}, with the darker and lighter bands representing statistical and systematic errors, respectively, covering the relevant data points.
 
\begin{figure}
    \centering
    \includegraphics[width=0.44\textwidth]{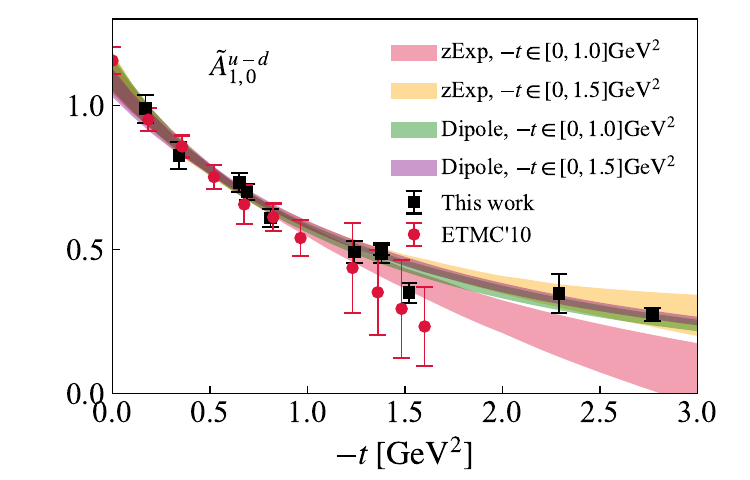}
    \includegraphics[width=0.44\textwidth]{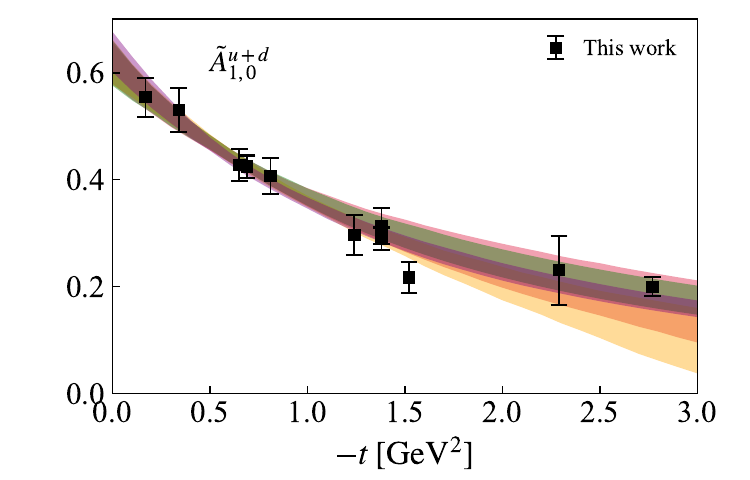}
    \includegraphics[width=0.44\textwidth]{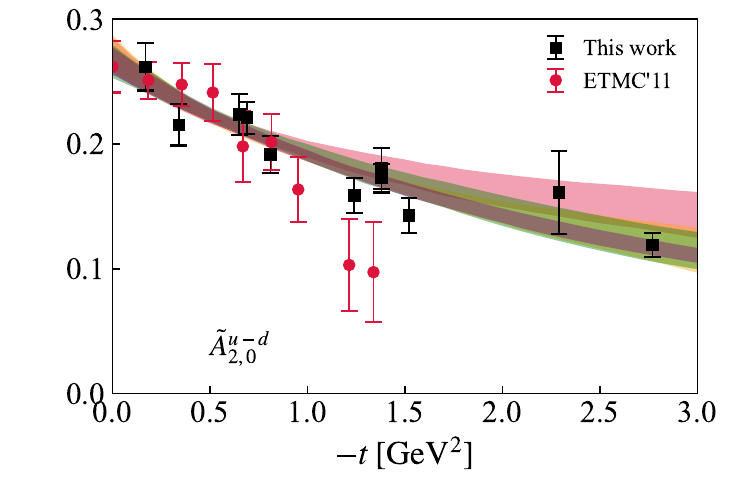}
    \includegraphics[width=0.44\textwidth]{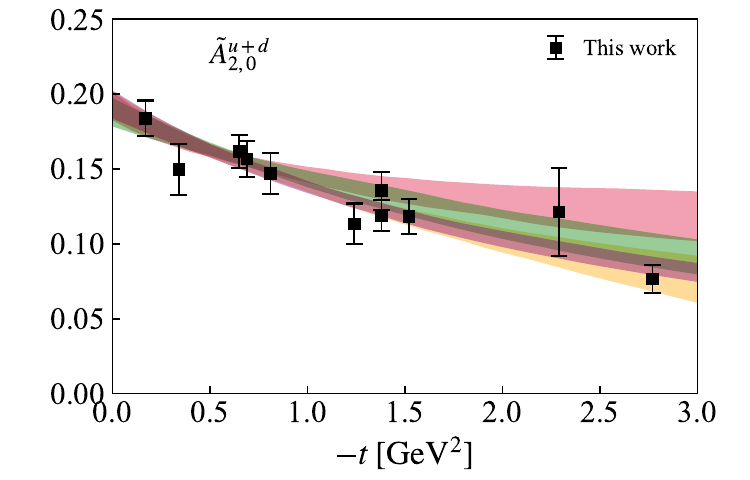}
    \caption{The first and second Mellin moments $\tilde{A}_{1,0}$ and $\tilde{A}_{2,0}$ of the axial-vector GPDs are shown. The left panels are the iso-vector results while the right panels are the iso-scalar results. The bands are results from $z$-expansion (zExp) and dipole methods by fitting data in the ranges $-t\in[0,1.0]~\rm{GeV}^2$ and $-t\in[0,1.5]~\rm{GeV}^2$. For comparison, we also show results derived from traditional local operator methods with the same pion mass for the iso-vector case (ETMC)~\cite{Alexandrou:2010hf,Baran:2011hq}. }
    \label{fig:A10A20}
\end{figure}

\begin{figure}
    \centering
    \includegraphics[width=0.44\textwidth]{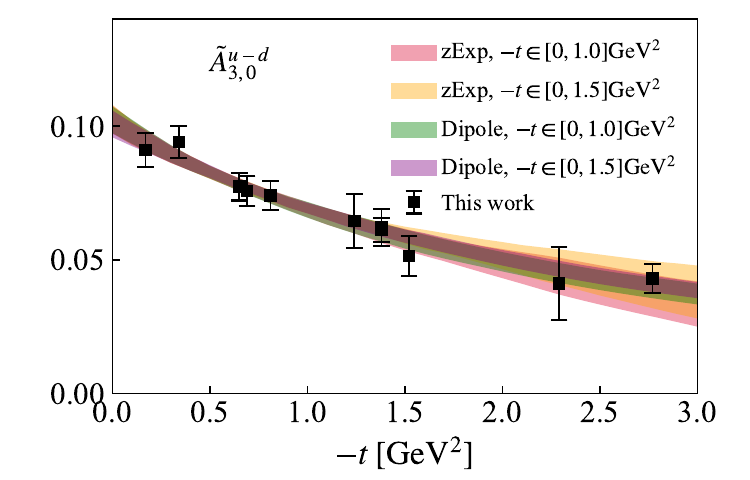}
    \includegraphics[width=0.44\textwidth]{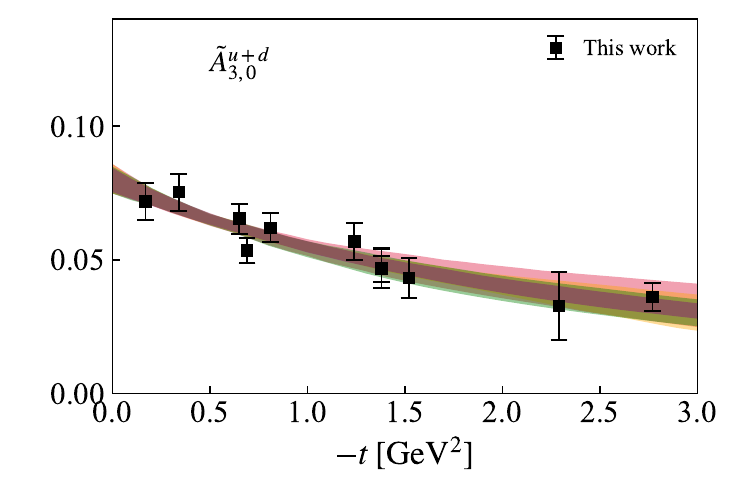}
    \includegraphics[width=0.44\textwidth]{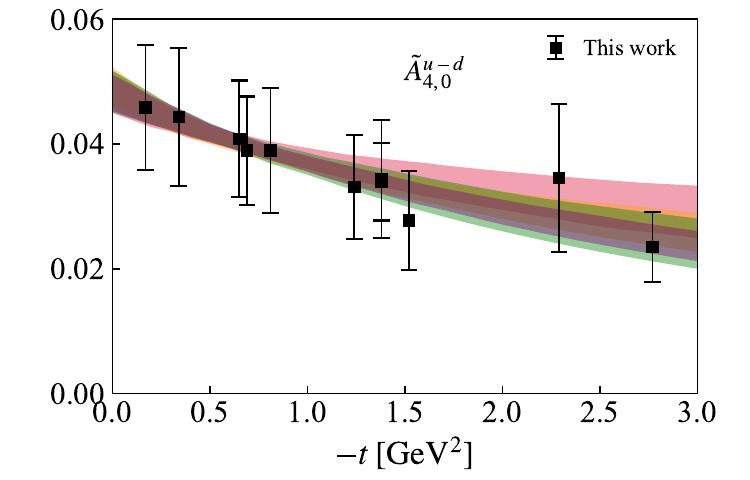}
    \includegraphics[width=0.44\textwidth]{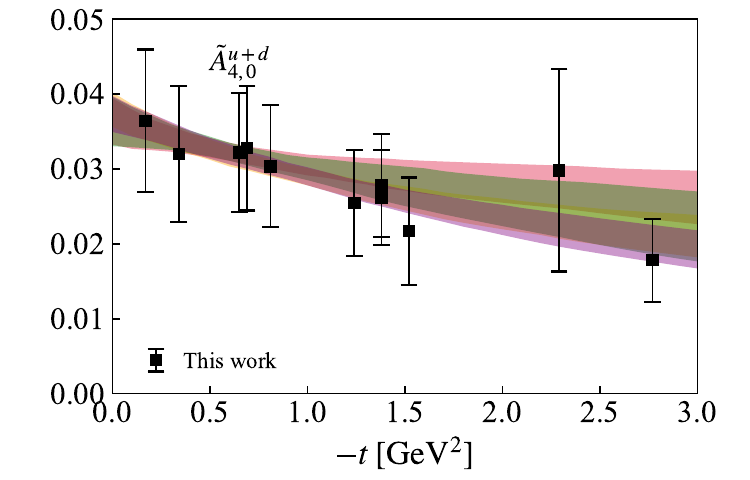}
    \includegraphics[width=0.44\textwidth]{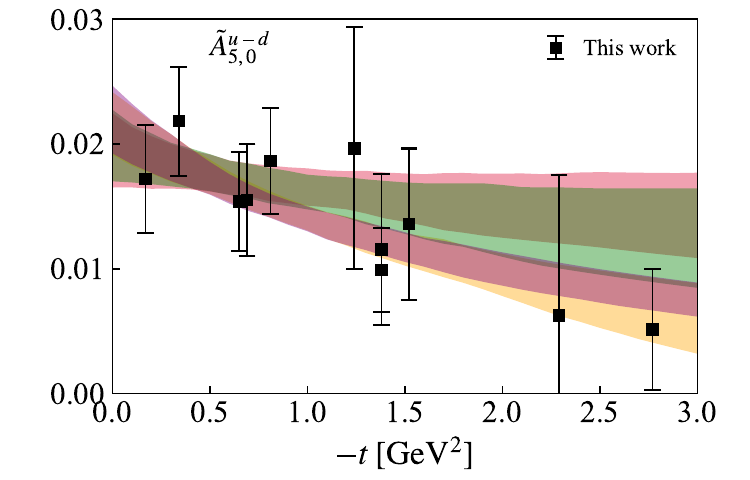}
    \includegraphics[width=0.44\textwidth]{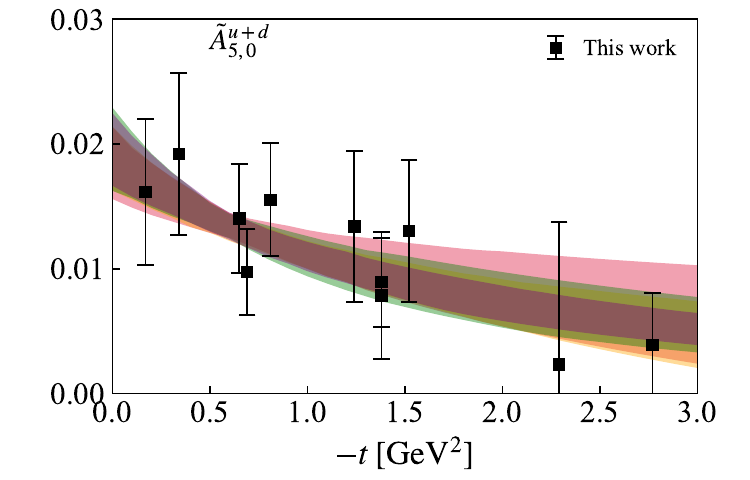}
	\caption{The Mellin moments $\tilde{A}_{3,0}$, $\tilde{A}_{4,0}$ and $\tilde{A}_{5,0}$ of the axial-vector GPDs are shown. The left panels are the iso-vector results while the right panels are the iso-scalar results. The bands are results from $z$-expansion (zExp) and dipole methods by fitting data in the ranges $-t\in[0,1.0]~\rm{GeV}^2$ and $-t\in[0,1.5]~\rm{GeV}^2$.\label{fig:A30A40A50}}
\end{figure}

We extended this analysis to all other values of the momentum transfer listed in \tb{stat}, covering both iso-vector and iso-scalar cases. For the latter, we neglected the mixing with the gluon distribution starting from $\mathcal{O}(\alpha_s)$~\cite{Yao:2022vtp}. In Fig.~\ref{fig:A10A20}, we summarize our determination of first two moments, $\tilde{A}_{1,0}$ and $\tilde{A}_{2,0}$, as functions of $-t$. For comparison, we show results obtained from traditional local operator methods with a similar lattice setup and pion mass for the iso-vector case (ETMC)~\cite{Alexandrou:2010hf,Baran:2011hq}. It is encouraging that our results align with the previous ETMC findings, suggesting that our extraction of the moments from non-local operators is effective. Notably, we are also able to extract higher moments up to $\tilde{A}_{5,0}$ for the first time, as shown in \fig{A30A40A50}. As one can see from the figure, we obtain a reasonable signal, and the $-t$ dependence of the results follows the general expectations. We note that we apply the $z$-expansion and a dipole fit, and for each we compare the parametrization of the $t$ dependence using data up to 1.0 GeV$^2$ and 1.5 GeV$^2$. In most cases, the dipole fit and the $z$-expansion are in agreement. However, a difference is found between $-t_{\rm max}=1.0$ GeV$^2$ and $-t_{\rm max}=1.5$ GeV$^2$. More details are given in Sec.\ref{sec:tdependence}.

\section{Insights into nucleon spin dynamics from axial vector GPD}\label{sec:spin}

GPDs offer crucial insights into the spin structure of the nucleon. For instance, in a longitudinally polarized nucleon, the contributions to the nucleon spin $\textbf{S}^N_z$ from quark helicity, orbital angular momentum (OAM), and total spin are represented by $\langle \textbf{S}^q_z\textbf{S}^N_z\rangle$, $\langle \textbf{L}^q_z\textbf{S}^N_z\rangle$ and $\langle \textbf{J}^q_z\textbf{S}^N_z\rangle$, respectively, where $\langle ...\rangle$ denotes the appropriate average. At $-t=0$, the first moment of the axial-vector GPD $\widetilde{H}$ is directly connected to the quark helicity contribution, expressed as, 
\begin{align}\label{eq:quarkheli}
    S^q_z=\frac{1}{2}\int_{-1}^1 dx \widetilde{H}^q(x,0,0)=\frac{1}{2}\tilde{A}^q_{1,0}(0).
\end{align}
According to Ji's spin sum rule~\cite{Ji:1996ek}, the contribution of the total angular momentum of quarks $J_z^q$ to the nucleon spin can be derived from the second moments of unpolarized quark GPDs, 
\begin{align}
    J^q_z=\frac{1}{2}(A^q_{2,0}(0)+B^q_{2,0}(0)),
\end{align}
where $A^q_{2,0}(0)=\int_{-1}^1dx x H^q(x,\xi=0,t=0)$ and $B^q_{2,0}(0)=\int_{-1}^1dx x E^q(x,\xi=0,t=0)$. We have determined $J^q_z$ using the same lattice setup and framework in \refcite{Bhattacharya:2023ays}. This allows us to determine the quark OAM contribution through,
\begin{align}
    L^q_z=J^q_z-S^q_z=\frac{1}{2}(A^q_{2,0}(0)+B^q_{2,0}(0))-\frac{1}{2}\tilde{A}^q_{1,0}(0)\,.
\end{align}
The quark spin-orbit correlation $\langle \textbf{S}^q_z\textbf{L}^q_z\rangle$ within the nucleon is another crucial aspect that can be probed through GPDs. This correlation describes how a quark's spin orientation is correlated to its orbital motion, offering deeper insights into the internal dynamics of hadrons. The spin-orbit correlation 
$C_z^q$ can be derived as follows~\cite{Lorce:2014mxa, Rajan:2017cpx, Rodekamp:2023wpe}\footnote{The quark spin-orbit correlation was first introduced using Wigner phase space distributions~\cite{Lorce:2011kd}; see also Ref.~\cite{Kanazawa:2014nha}. It depends on the path chosen for the Wilson line in the definition of the Wigner functions.  The definition we are using here corresponds to a straight Wilson line connecting the quark fields of the corresponding bi-local operator.
Related work can also be found in~\cite{Bhattacharya:2024sno, Bhattacharya:2024sck, Hatta:2024otc} and references therein.},
\begin{align}\label{eq:spinorbit}
\begin{split}
    C_z^q&=\frac{1}{2}\int_{-1}^1 dx x \widetilde{H}^q(x,0,0)-\frac{1}{2}\left[\int_{-1}^1 dx H^q(x,0,0)-\frac{m_q}{2m_N}\int_{-1}^1dx(E^q_T(x,0,0)+2\widetilde{H}^q_T(x,0,0))\right]\\
    &\approx \frac{1}{2}\int_{-1}^1 dx x \widetilde{H}^q(x,0,0)-\frac{1}{2}\int_{-1}^1 dx H^q(x,0,0)\\
    &=\frac{1}{2}(\tilde{A}^q_{2,0}(0)-A^q_{1,0}(0)),
\end{split}
\end{align}
where we have ignored the term suppressed by the light quark mass. To determine $C_z^q$, one needs additional input from the first moment  of the unpolarized quark GPD, $A^q_{1,0}(0)=\int_{-1}^1dx H^q(x,\xi=0,t=0)$, giving the total quark number inside the nucleon.

In this section, we will discuss the quark helicity $S^q_z$, OAM $L^q_z$ and spin-orbit correlation $C^q_z$ derived from our GPD moments. With the broad range of $-t$ values available, we can also perform a Fourier transform and explore their distribution in impact parameter space.

\subsection{Quark helicity, OAM, and spin-orbit correlations in nucleons}\label{sec:tdependence}

In \sec{moments}, we have extracted the Mellin moments of axial-vector GPD $\widetilde{H}(x,\xi=0,-t)$ up to the fifth order. With multiple values of $-t$ ranging from 0.17 to 2.77 $\rm GeV^2$, we can parameterize the $-t$ dependence and extrapolate to $-t\rightarrow0$. One commonly used model in fitting nucleon form factors and moments is the dipole model, 
\begin{align}\label{eq:dipole}
    \tilde{A}_{n,0}(-t)=\frac{\tilde{A}_{n,0}(0)}{(1-\frac{t}{M^2})^2},
\end{align}
where $\tilde{A}_{n,0}(0)$ and $M$ are fit parameters. Empirically, this model has been successful in fitting form factors from experiments and lattice QCD at low $-t$. However, when the data span a wide range of $-t$,  a more flexible parameterization is often preferred, such as the $z$-expansion series~\cite{Lee:2015jqa},
\begin{align}\label{eq:zexp}
    \tilde{A}_{n,0}(-t)=\sum_{k=0}^{k_{\rm max}}a_kz(t)^k,
\end{align}
with,
\begin{align}
    z(t)=\frac{\sqrt{t_{\rm cut}-t}-\sqrt{t_{\rm cut}-t_0}}{\sqrt{t_{\rm cut}-t}+\sqrt{t_{\rm cut}-t_0}}
\end{align}
where $t_0$ is selected to minimize the span of $z(t)$ over the given range of $t$, thereby optimizing the convergence of the series expansion. In this work, we apply 
$t_0=t_{\rm cut}(1-\sqrt{1-t_{\rm max}/t_{\rm cut}})$ 
with $t_{\rm cut}$ set to be the three-pion kinematic threshold $(3m_\pi)^2$. 
To maintain a reasonable $\chi^2/{\rm d.o.f.}$ and avoid overfitting, we truncate the series at $k_{\rm max}=2$ in this work. To stabilize the fit, we imposed a Gaussian prior to the $|a_k/a_0|$ with a central value of 0 and width $|a_k/a_0|_{\rm max}$ = 5.

To estimate the model bias, we vary the range of $-t$ included in the fit, specifically $-t\in[0,1.0]~\rm{GeV}^2$ and $-t\in[0,1.5]~\rm{GeV}^2$. Interestingly, the results from the dipole model not only go through the data point included in the fit but also can describe the data points at extended region up to 3 $\rm GeV^2$. In contrast, the $z$-expansion model, although flexible, sometimes fails to describe data points at higher $-t$ values, as it is a series expansion that can become unstable when extrapolated too far. For our final estimates, we average the results from different model choices and $-t$ ranges, taking their deviations as systematic errors, as done in previous studies~\cite{Gao:2020ito,Bhattacharya:2023ays}. Our estimates of $\tilde{A}_{n+1,0}$ at $-t=0$ are summarized in \tb{fitQsqx01} for both the iso-vector and iso-scalar cases. The statistical and systematic uncertainties, shown in the first and second round brackets, respectively, are estimated based on the mean and deviation of results across various model choices and $-t$ ranges, following the approach used in previous studies~\cite{Gao:2020ito,Bhattacharya:2023ays}. Our determination of the first two moments for the iso-vector case agrees with results from traditional local operators with a similar lattice setup and pion mass from ETMC~\cite{Alexandrou:2010hf, Baran:2011hq}, where $\tilde{A}_{1,0}^{u-d}$=1.156(47) and $\tilde{A}_{2,0}^{u-d}$=0.262(21). We repeat that for the iso-scalar moments we omitted disconnected diagrams,
which were found to be small on this ensemble~\cite{Alexandrou:2021oih}, and the mixing with gluons in the perturbative matching, which starts at $\mathcal{O}(\alpha_s)$~\cite{Yao:2022vtp}.

\begin{table}[h!]
\centering
\begin{tabular}{c c c c}
\hline
\hline
$\tilde{A}_{1,0}^{u-d}$ &1.110(57)(19)&$\tilde{A}_{1,0}^{u+d}$ &0.625(38)(12)  \cr
\hline
$\tilde{A}_{2,0}^{u-d}$ &0.270(12)(05)&$\tilde{A}_{2,0}^{u+d}$ &0.191(9)(3)  \cr
\hline
$\tilde{A}_{3,0}^{u-d}$ &0.102(5)(1)&$\tilde{A}_{3,0}^{u+d}$ &0.080(5)(1)  \cr
\hline
$\tilde{A}_{4,0}^{u-d}$ &0.049(3)(1)&$\tilde{A}_{4,0}^{u+d}$ &0.037(3)(1)  \cr
\hline
$\tilde{A}_{5,0}^{u-d}$ &0.021(3)(1)&$\tilde{A}_{5,0}^{u+d}$ &0.019(3)(1) \cr
\hline
\hline
\end{tabular}
\caption{The Mellin moments of the axial-vector GPDs, $\widetilde{H}$, extrapolated to $-t = 0$ are presented. The statistical and systematic uncertainties, shown in the first and second round brackets, respectively, are estimated based on the mean and deviation of results across various model choices and $-t$ ranges, following the approach used in previous studies~\cite{Gao:2020ito,Bhattacharya:2023ays}.}
\label{tb:fitQsqx01}
\end{table}

The quark helicity can be derived from the first moment according to \Eq{quarkheli}, yielding,
\begin{align}
    S^{u-d}_z=0.555(29)(9),\qquad S^{u+d}_z=0.313(19)(6).
\end{align}
Our findings indicate that the light quark ($u+d$) helicity contribute significantly to the nucleon spin, consistent with results from previous lattice calculations (see, e.g., a recent review in ~\refcite{Liu:2021lke}). In \refcite{Bhattacharya:2023ays}, using the same lattice setup and methodology, we determined the total quark contribution to the nucleon spin as $J^{u-d}_z=0.281(21)(11)$ and $J^{u+d}_z=0.296(22)(33)$. Consequently, we infer the quark orbital angular momentum (OAM) as,
\begin{align}\label{eq:Lz}
    L^{u-d}_z=-0.260(34)(19),\qquad L^{u+d}_z=-0.010(37)(8).
\end{align}
Interestingly, this result suggests that the light quark ($u+d$) OAM is very small~\cite{Liu:2021lke}.
However, this does not imply that the OAM of individual quarks are negligible, as the finite value of $L^{u-d}_z$ indicates. Instead, the small total OAM for light quarks are likely due to the opposing signs of the OAM contributions from different quark flavors, leading to a cancellation effect.

Finally, we estimate the quark spin-orbit correlation using \Eq{spinorbit}, combining $\tilde{A}_{2,0}$ from this work with $A_{1,0}$ from \refcite{Bhattacharya:2023ays}, which gives,
\begin{align}\label{eq:Cz}
    C^{u-d}_z=-0.356(20)(5),\qquad C^{u+d}_z=-1.325(39)(14).
\end{align}
These results are in close agreement with previous findings in \refcites{Lorce:2014mxa, Rodekamp:2023wpe}. Additionally, it is observed that the iso-vector spin-orbit correlation is significantly smaller than the iso-scalar one, consistent with predictions from the large $N_c$ limit~\cite{Kim:2024cbq}, which suggests that $C^{u-d}_z=\mathcal{O}(N_c^0)$ and $C^{u+d}_z=\mathcal{O}(N_c^1)$. 

We note that, although these outcomes are encouraging, it is crucial to address systematic uncertainties arising from unphysical quark masses, disconnected diagrams, lattice discretization errors, and excited state contaminations in future works to achieve higher precision.

\subsection{Impact-parameter-space interpretation}

\begin{figure}
    \centering
    \includegraphics[width=0.45\textwidth]{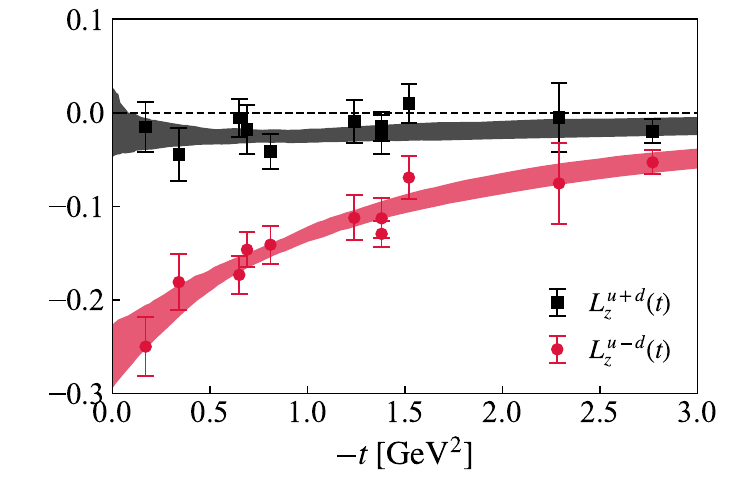}
    \includegraphics[width=0.45\textwidth]{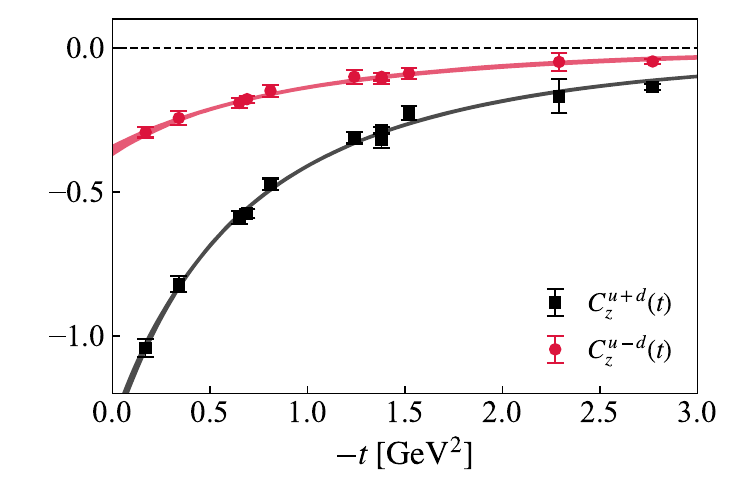}
	\caption{$L^{q}_z(-t)$ (left panel) and $C^{q}_z(-t)$ (right panel) are shown as a function of $-t$ for both the iso-vector and iso-scalar cases. The bands are derived from the dipole model fitted from the moments $\tilde{A}_{n,0}(-t)$.\label{fig:LzCz}}
\end{figure}

\begin{figure}
    \centering
    \includegraphics[width=0.9\textwidth]{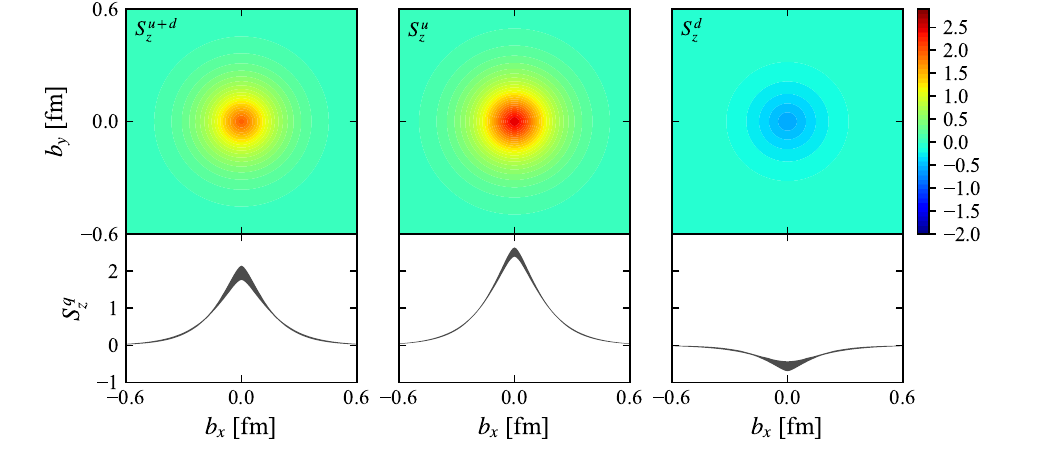}
	\caption{Upper panels: the light quark helicity density in the impact-parameter plane. Lower panels: the light quark helicity density as a function of $b_x$ with $b_y=0$, including its uncertainties. \label{fig:DensitySz}}
\end{figure}

\begin{figure}
    \centering
    \includegraphics[width=0.9\textwidth]{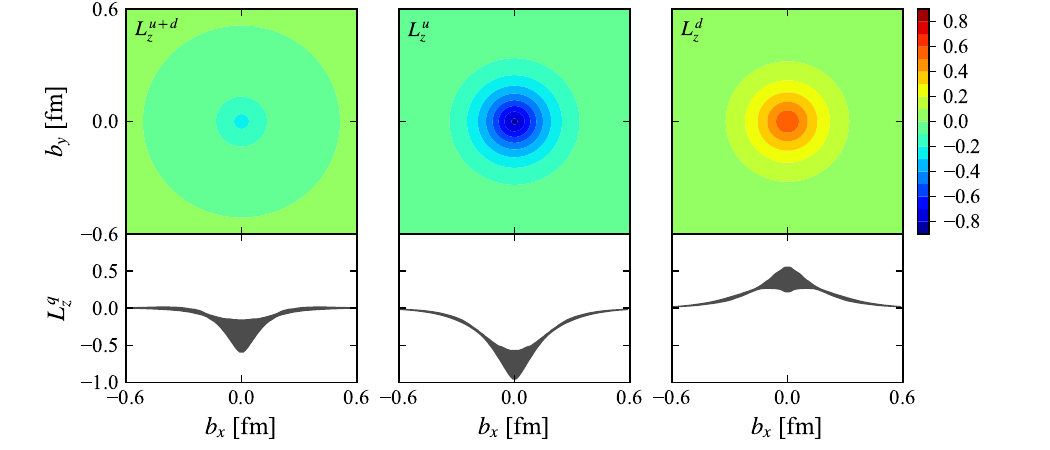}
	\caption{Upper panels: the light quark OAM density in the impact-parameter plane. Lower panels: the light quark OAM density as a function of $b_x$ with $b_y=0$, including its uncertainties.\label{fig:DensityLz}}
\end{figure}

\begin{figure}
    \centering
    \includegraphics[width=0.9\textwidth]{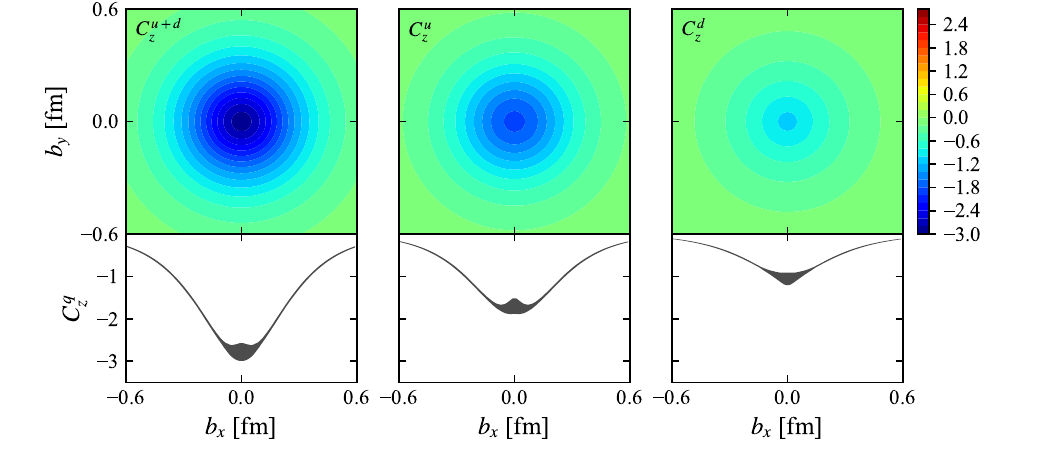}
	\caption{Upper panels: the light quark spin-orbit correlation density in the impact-parameter plane. Lower panels: the light quark spin-orbit correlation density as a function of $b_x$ with $b_y=0$, including its uncertainties.\label{fig:DensityCz}}
\end{figure}

In addition to the spin decomposition discussed above, the GPDs also provide crucial insights into the three-dimensional structure of the nucleon. By performing a Fourier transform (FT) over the momentum transfer $-t$, we can derive parton distributions in the impact-parameter space. For instance, the quark helicity distribution in this space is given by, 
\begin{equation}
    \tilde{q}(x, \mathbf{b}_\perp) = \int \frac{{\rm d}^2 \mathbf{\Delta}_\perp}{(2\pi)^2} \widetilde{H}(x, \xi=0, \mathbf{\Delta}^2_\perp) e^{i\mathbf{b}_\perp\cdot\mathbf{\Delta}_\perp},
\end{equation}
where $-t=\mathbf{\Delta}^2_\perp$. Its moments $\int_{-1}^1 dx x^n\tilde{q}(x, \mathbf{b}_\perp)$, which are the FT of the $\tilde{A}_{n+1,0}(-t)$, are averaged distributions with weight $x^n$ in the impact-parameter plane. In \fig{A10A20}, we have shown the $\tilde{A}_{1,0}(-t)=2S^q_z(-t)$. Similar to \Eq{Lz} and \Eq{Cz}, the quark OAM $L^q_z$ and spin-orbit correlation $C^q_z$ can also be defined with respect to momentum transfer $-t$. The results are presented in \fig{LzCz} for both the iso-vector and iso-scalar cases. The bands are derived using the dipole model fit from \sec{tdependence} by taking the average of the results obtained from the two aforementioned fit ranges. As shown in the figures, the total light-quark OAM, $L_z^{u+d}$, remains small, and the magnitudes of both $L^q_z(-t)$ and $C^q_z(-t)$ decrease rapidly as $-t$ increases. For the Fourier transform over $-t$, we rely on the fit results from the dipole model, as no data are available beyond 3 $\rm GeV^2$. The results at small $b_T$ may be affected by systematic errors due to the model assumptions, though these errors are likely minimal since the moments decay quickly at large $-t$.  

In \fig{DensitySz}, the quark helicity density in the impact-parameter plane is depicted. As one can see, the $u$ and $d$ quark densities have opposite signs, but the $u$ quark’s magnitude is significantly larger, resulting in a positive combined helicity contribution from $u + d$ quarks.

The quark OAM density in the impact-parameter plane, shown in \fig{DensityLz}, also exhibits opposite signs for $u$ and $d$ quarks. This suggests that $u$ and $d$ quarks orbit the longitudinal momentum in opposite directions. Notably, since the magnitudes of $u$ and $d$ quarks are similar, their sum is nearly zero. This observation is very different from the quark number density distribution we studied in \refcite{Bhattacharya:2023ays}, where the $d$ quark shows lower magnitudes in the impact-parameter space compared to the $u$ quark. This difference may imply that $d$ quarks possess a larger $k_T$ for a given $b_T$. 

Lastly, for the first time, we present results for the quark spin-orbit correlation distribution in impact-parameter space. We find that these correlations are negative for both light quarks, with the $u$ quark exhibiting a larger magnitude. Interestingly, the sign of $C_z^q = \langle \textbf{S}_z^q \textbf{L}_z^q \rangle$ matches that of $S_z^q \cdot L_z^q = \langle \textbf{S}_z^q \textbf{S}_z^N \rangle \cdot \langle \textbf{L}_z^q \textbf{S}_z^N \rangle$ for each quark flavor. Additionally, unlike $S^q$ and $L^q$, which approach zero rapidly as $b_T$ nears the nucleon size, there is still a non-zero value observed for the spin-orbit correlation.

\section{Conclusion}\label{sec:conclusion}

In this work, we present a study of the moments of axial-vector GPD $\widetilde{H}$ using lattice QCD. We compute the quasi-GPD matrix elements in an asymmetric frame with multiple values of the momentum transfer, allowing us to study the $t$ dependence.
The quasi-GPDs matrix elements are then renormalized using the ratio scheme. We employ the short-distance factorization framework to extract the first few moments of the GPDs. For the first time, we obtain results for up to and including the fifth moment of axial-vector GPD $\widetilde{H}$ with reasonable signal and $t$ dependence as summarized in \tb{allisov} and \tb{allisos}. Our determination of the first two moments is consistent with previous calculations using traditional local operator methods. 
From these moments we infer the quark helicity and OAM contributions to the nucleon spin as well as the quark spin-orbit correlations. In agreement with previous findings, our results indicate that the light quark helicity contributes significantly to the nucleon spin, while the OAM of individual quark flavors shows an interesting pattern of cancellation, leading to a small net OAM for the light quarks. Additionally, the spin-orbit correlations for both light quarks are found to be negative, aligning with the sign of $S^q_z \cdot L^q_z$. The magnitudes of the iso-vector and iso-scalar combinations are in agreement with the large-$N_c$ predictions. We emphasize that the use of an asymmetric frame with multiple values of $t$ enables us to explore impact-parameter space distributions via a Fourier transform over $t$. That provides us multiple images of the nucleon spin structure, showcasing the spatial distributions of quark helicity, orbital angular momentum (OAM) and, for the first time, the spin-orbit correlations in the transverse plane. These distributions exhibit distinct features for different quark flavors, offering a deeper understanding of the nucleon’s internal structure. However, we acknowledge several systematic uncertainties that were not addressed in this exploratory work. Future research will aim to control these uncertainties, including contributions from disconnected diagrams and gluon mixing in the iso-scalar case. It will also involve a careful analysis of excited-state contamination with multiple source-sink separations, calculations with quark masses at the physical point, and the use of multiple lattice spacings to achieve the continuum limit. These efforts aim to further refine our understanding of nucleon spin dynamics.

\begin{table}[t]
\centering
\begin{tabular}{c c c c c c}
\hline
$-t~\rm{GeV}^2$&$\tilde{A}_{10}^{u-d}$&$\tilde{A}_{20}^{u-d}$&$\tilde{A}_{30}^{u-d}$&$\tilde{A}_{40}^{u-d}$&$\tilde{A}_{50}^{u-d}$\cr
\hline
0.17 & 0.990(47)(01) & 0.262(14)(05) & 0.091(05)(01) & 0.046(03)(07) & 0.017(02)(02) \cr
0.34 & 0.827(45)(01) & 0.215(11)(06) & 0.094(05)(01) & 0.044(03)(08) & 0.022(03)(02) \cr
0.65 & 0.733(32)(01) & 0.223(11)(05) & 0.077(04)(01) & 0.041(02)(07) & 0.015(02)(02) \cr
0.69 & 0.699(27)(02) & 0.221(08)(05) & 0.076(04)(02) & 0.039(02)(06) & 0.016(02)(03) \cr
0.81 & 0.609(31)(01) & 0.192(10)(05) & 0.074(04)(01) & 0.039(03)(07) & 0.019(02)(02) \cr
1.24 & 0.491(33)(04) & 0.159(10)(04) & 0.065(06)(04) & 0.033(03)(06) & 0.020(03)(06) \cr
1.38 & 0.485(29)(02) & 0.181(13)(04) & 0.062(05)(02) & 0.034(04)(05) & 0.012(03)(03) \cr
1.38 & 0.502(20)(01) & 0.173(09)(03) & 0.061(03)(01) & 0.034(02)(04) & 0.010(02)(02) \cr
1.52 & 0.351(32)(02) & 0.143(10)(04) & 0.051(05)(02) & 0.028(03)(05) & 0.014(03)(03) \cr
2.29 & 0.348(63)(04) & 0.161(30)(04) & 0.041(10)(04) & 0.035(07)(05) & 0.006(06)(05) \cr
2.77 & 0.274(21)(02) & 0.119(07)(02) & 0.043(04)(02) & 0.023(02)(03) & 0.006(05)(02) \cr
\hline
\end{tabular}
\caption{The table of iso-vector moments $\tilde{A}_{n+1,0}^{u-d}$.
}
\label{tb:allisov}
\end{table}

\begin{table}[t]
\centering
\begin{tabular}{c c c c c c}
\hline
$-t~\rm{GeV}^2$&$\tilde{A}_{10}^{u+d}$&$\tilde{A}_{20}^{u+d}$&$\tilde{A}_{30}^{u+d}$&$\tilde{A}_{40}^{u+d}$&$\tilde{A}_{50}^{u+d}$\cr
\hline
0.17 & 0.554(34)(03) & 0.184(07)(05) & 0.072(04)(03) & 0.036(03)(06) & 0.016(02)(03) \cr
0.34 & 0.531(40)(02) & 0.150(13)(04) & 0.075(05)(02) & 0.032(03)(06) & 0.020(03)(03) \cr
0.65 & 0.427(29)(02) & 0.162(07)(04) & 0.065(04)(02) & 0.032(02)(06) & 0.014(02)(02) \cr
0.69 & 0.424(20)(01) & 0.157(07)(05) & 0.053(04)(01) & 0.033(02)(06) & 0.010(02)(01) \cr
0.81 & 0.407(33)(01) & 0.147(10)(04) & 0.062(04)(01) & 0.030(02)(06) & 0.016(03)(02) \cr
1.24 & 0.296(35)(02) & 0.113(11)(03) & 0.057(05)(02) & 0.025(04)(04) & 0.013(03)(03) \cr
1.38 & 0.313(32)(02) & 0.135(10)(03) & 0.047(06)(02) & 0.028(03)(04) & 0.008(03)(02) \cr
1.38 & 0.290(20)(01) & 0.119(07)(03) & 0.047(04)(01) & 0.026(02)(04) & 0.009(02)(02) \cr
1.52 & 0.217(28)(02) & 0.118(09)(03) & 0.043(06)(02) & 0.022(03)(04) & 0.013(03)(02) \cr
2.29 & 0.231(61)(04) & 0.121(24)(05) & 0.033(09)(04) & 0.030(07)(07) & 0.002(07)(05) \cr
2.77 & 0.199(16)(02) & 0.077(07)(03) & 0.036(04)(02) & 0.018(02)(03) & 0.004(02)(02) \cr
\hline
\end{tabular}
\caption{The table of iso-scalar moments $\tilde{A}_{n+1,0}^{u+d}$.
}
\label{tb:allisos}
\end{table}

\section*{Acknowledgement}

This material is based upon work supported by the U.S. Department of Energy, Office of Science, Office of Nuclear Physics through Contract No.~DE-SC0012704, No.~DE-AC02-06CH11357 and within the framework of Scientific Discovery through Advance Computing (SciDAC) award Fundamental Nuclear Physics at the Exascale and Beyond. 
The work of S.~B. has been supported by the Laboratory Directed Research and Development program of Los Alamos National Laboratory under project number 20240738PRD1.
S.~B. has also received support from the U.~S. Department of Energy through the Los Alamos National Laboratory. Los Alamos National Laboratory is operated by Triad National Security, LLC, for the National Nuclear Security Administration of U.~S. Department of Energy (Contract No. 89233218CNA000001). 
K.~C.\ is supported by the National Science Centre (Poland) grants SONATA BIS no.\ 2016/22/E/ST2/00013 and OPUS no.\ 2021/43/B/ST2/00497. 
M.~C. and J. M. acknowledge financial support by the U.S. Department of Energy, Office of Nuclear Physics, Early Career Award under Grant No.\ DE-SC0020405, as well as Grant No.\ DE-SC0025218.
The work of A.~M. is supported by the National Science Foundation under grant number PHY-2412792.
F.~S.\ was funded by the NSFC and the Deutsche Forschungsgemeinschaft (DFG, German Research Foundation) through the funds provided to the Sino-German Collaborative Research Center TRR110 “Symmetries and the Emergence of Structure in QCD” (NSFC Grant No. 12070131001, DFG Project-ID 196253076 - TRR 110). 
The authors also acknowledge partial support by the U.S. Department of Energy, Office of Science, Office of Nuclear Physics under the umbrella of the Quark-Gluon Tomography (QGT) Topical Collaboration with Award DE-SC0023646.
Computations for this work were carried out in part on facilities of the USQCD Collaboration, which are funded by the Office of Science of the U.S. Department of Energy. 
This research used resources of the Oak Ridge Leadership Computing Facility, which is a DOE Office of Science User Facility supported under Contract DE-AC05-00OR22725.
This research used resources of the National Energy Research
Scientific Computing Center, a DOE Office of Science User Facility
supported by the Office of Science of the U.S. Department of Energy
under Contract No. DE-AC02-05CH11231 using NERSC award
NP-ERCAP0027642.
This research was supported in part by PLGrid Infrastructure (Prometheus supercomputer at AGH Cyfronet in Cracow).
Computations were also partially performed at the Poznan Supercomputing and Networking Center (Eagle supercomputer), the Interdisciplinary Centre for Mathematical and Computational Modelling of the Warsaw University (Okeanos supercomputer), and at the Academic Computer Centre in Gda\'nsk (Tryton supercomputer).
The gauge configurations have been generated by the Extended Twisted Mass Collaboration on the KNL (A2) Partition of Marconi at CINECA, through the Prace project Pra13\_3304 ``SIMPHYS".
Inversions were performed using the DD-$\alpha$AMG solver~\cite{Frommer:2013fsa} with twisted mass support~\cite{Alexandrou:2016izb}. 


\providecommand{\href}[2]{#2}\begingroup\raggedright\endgroup

\end{document}